\documentclass[12pt]{article}

\usepackage{amsmath,amssymb,graphicx,jheppub,comment}

	\title{Super-Giants in Gutowski-Reall Black Hole}
	\author{Nayan Mondal and Nemani V. Suryanarayana}
\affiliation[]{The Institute of Mathematical Sciences,  Taramani, Chennai 600113, India\\ \&}
\affiliation[]{Homi Bhabha National Institute, Anushakti Nagar, Mumbai 400094, India.}
	\emailAdd{nayanm, nemani@imsc.res.in}

	\abstract{We present all bosonic giant and dual-giant type configurations of a probe D3-brane in the BPS single-parameter Gutowski-Reall black hole in 10d type IIB supergravity that do not break any of its supersymmetries. The resulting D3-brane world-volumes can be given by the common zeros of three holomorphic functions of five complex scalar harmonics of the geometry. These probe branes support world-volume electromagnetic fields which we characterise completely in terms of pull-backs of closed 2-forms. Our configurations can be seen as natural generalisations of known supersymmetric D3-branes in $AdS_5 \times S^5$ and approach them far away from the black hole horizon.}

\begin{document}
\maketitle
	
\vskip .5cm
\section{Introduction}

There exist supersymmetric black hole solutions of type IIB supergravity that asymptote to $AdS_5 \times S^5$ and preserve just two supersymmetries. First of these was a 1-parameter supersymemtric $AdS_5$ black hole solution found in minimal ${\cal N}=1$ gauged supergravity by Gutowski and Reall in \cite{Gutowski:2004ez,Gutowski:2004yv}. It was shown in \cite{Gauntlett:2004cm} that this when lifted to a solution of type IIB it is 1/16-BPS. These Gutowski-Reall (GR) black holes  have two equal angular momenta and three equal R-charges. Over the years there have been many generalisations \cite{Cvetic:2004hs, Cvetic:2004ny, Cvetic:2005zi, Chong:2005hr, Wu:2011gq}
 both with and without supersymmetries. 

From the holographic dual boundary theory side the study of the corresponding 1/16-BPS states in the ${\cal N}=4$, $SU(N)$ SYM theory on $S^3 \times {\mathbb R}$ spacetime has been pursued in various works (see, for instance, \cite{Grant:2008sk, Chang:2013fba, Yokoyama:2014qwa, SARPNVS}). The states in this sector that have enhanced supersymmetries are not expected to have bulk duals that admit horizons. In fact in the 1/2-BPS sector all the duals are smooth geometries as described by LLM geometries \cite{Lin:2004nb} (or null-sigularities like the superstar). Those 1/16-BPS states that have 4 or 8 supersymmetries are also expected to be dual to only smooth geometries. However, the states with just two supersymmetries (1/16-BPS states), on general grounds, are expected to be dual to geometries with no-horizons (smooth fuzz-balls), geometries with single horizon etc, and to account for all the duals of 1/16-BPS states of the gauge theory one needs to take all these into account. Therefore, one expects there to be classes of 1/16-BPS smooth horizon-less geometries, and one such geometry was studied in \cite{Gauntlett:2004cm} -- referred to as the deformed $AdS_5 \times S^5$. Inspite of all the progress with the construction of 1/16-BPS black holes in $AdS_5 \times S^5$, the most general single-horizon geometries are not yet fully known.\footnote{See, for instance,  \cite{Sinha:2006sh, Sinha:2007ni}, for BPS D3-branes in the near horizon geometries of the GR black holes.} In fact it has been conjectured by Minwalla et al \cite{Bhattacharyya:2010yg, Choi:2024xnv} that most general 1/16-BPS black hole may admit hair that does not break its supersymmetries. These hair may be given by the back reaction of supersymmetric D3-branes in the black hole geometries, that do not destroy the horizon. This makes it important to construct finite energy probe D3-branes in these black holes. Some such probe D3-branes are already known (see \cite{Aharony:2021zkr}, for instance) - particularly in the background of the original Gutowski-Reall black hole. 

In the $AdS_5 \times S^5$ background the BPS probe D3-branes have been known for a long time. They include the Mikhailov giants \cite{Mikhailov:2000ya}, the wobbling dual-giants \cite{Ashok:2008fa} and more generally the Kim-Lee configurations \cite{Kim:2006he}. In \cite{Ashok:2010jv} (see also, \cite{Kim:2005mw, Sinha:2007ni}) a description of all the BPS world-volume electromagnetic fields on any of the above giant gravitons is provided. The analogs of these solutions in the background of the 1/16-BPS geometries, though expected to exist, are not completely known. Finding these probes is expected to play interesting role in addressing various physics questions related to these geometries. Therefore, in this note, we address this limited question of finding all bosonic probe D3-brane configurations (that include the world-volume electromagnetic fields) that preserve both the supersymemtries of the 1/16-BPS type IIB geometries, namely the GR black hole (and some generalisations) and the smooth, horizon-free deformed $AdS_5 \times S^5$ of \cite{Gauntlett:2004cm}. We find that the D3-brane configurations that preserve the supersymmetries of the deformed $AdS_5 \times S^5$ are given by the same conditions as those with no deformation (pure $AdS_5 \times S^5$).
On the other hand, we show that, in the context of black holes there is a very rich class of these objects; of which the ones known earlier form a special sub-class. Our description of these general D3-brane giants is a non-trivial generalisation of the Kim-Lee description of 1/16-BPS giant gravitons in the $AdS_5 \times S^5$ geometry, where the holomorphic functions involved depend on five particular complex scalar harmonics of the black hole. We also provide a description of all the EM fields on these BPS D3-brane probes. 

The rest of this note is organised as follows. We set up the $\kappa$-projection conditions in section \ref{kappaanalysis} for a probe D3-brane in the Gutowski-Reall black hole. In section \ref{dualgiantsection} we solve the relevant BPS equations for wobbling dual-giants in terms of complex embedding functions. In section \ref{giantsection} we solve for Mikhailov type giants in the GR background. We find the Kim-Lee type description of the results encompassing those of sections \ref{dualgiantsection} and \ref{giantsection} in terms of three holomorphic functions in section \ref{kimleesection}. The problem of turning on EM fields is addressed in section \ref{emwavesection}. We conclude with a discussion of the results and open questions in section \ref{discussion}. The four appendices contain some additional results not covered in the main text.

\section{BPS D3-branes in GR black hole}
\label{kappaanalysis}
 \def\mathbi#1{\textbf{\em #1}} 
The Gutowski-Reall back hole \cite{Gutowski:2004ez,Gutowski:2004yv} can be lifted to a solution of the 10d type IIB  supergravity and it represents a supersymmetric black hole geometry, supported by a self-dual RR 5-form $F^{(5)}$, in $AdS_5 \times S^5$ preserving two of the 32 supersymmetries. The Killing spinor of this geometry was written down first in \cite{Gauntlett:2004cm}. Following the conventions of \cite{Gauntlett:2004cm} (with $\eta = 1$ there) the funfbein for $AdS_5$ part of this black hole are given by 
\begin{eqnarray}
      e^0 &=& (1-\frac{\omega^2}{r^2})[dt - \frac{r^2}{2 l}(1+\frac{2\omega^2}{r^2}+\frac{3\omega^2}{2 r^2 (r^2 - \omega^2)})\sigma_3^L],\nonumber\\
    e^1&=& \frac{l dr}{(1-\frac{\omega^2}{r^2})\sqrt{l^2+ r^2+2\omega^2}},\nonumber\\
    e^2 &=& \frac{r}{2} \sigma_1^L,\nonumber\\
    e^3 &=& \frac{r}{2}\sigma_2^L,\nonumber\\
    e^4 &=& \frac{r}{2l}\sqrt{l^2+r^2+2\omega^2} \sigma_3^L.\label{gr-ads5}
\end{eqnarray}
Here $\omega$ is a constant representing the location of the event horizon of the black hole and $l $ is the radius of $AdS_5$. Furthermore $\sigma_1^L, \; \sigma_2^L,\; \sigma_3^L$ are the following $SU(2)$ left-invariant one forms\begin{eqnarray}
       \sigma_1^L &=& \sin\phi \; d\theta - \sin\theta\cos\phi \, d\psi,\nonumber\\
       \sigma_2^L &=& \cos\phi \, d\theta+ \sin\theta\sin\phi \, d\psi,\nonumber\\
       \sigma_3^L &=& d\phi +\cos\theta \, d \psi.
   \end{eqnarray} 
   The coordinates $(t, r, \theta, \phi, \psi)$ have the following ranges : $ -\infty < t < \infty,$ $ 0\le r < \infty,$ $0\le \theta \le \pi$, $0\le \psi <2\pi$ and $0\le \phi <4\pi$.\footnote{Note that when $\omega = 0$ this geometry locally becomes that of global $AdS_5$ in the standard coordinates under the identification $\phi \rightarrow \phi-2t/l$.} The frame for the $S^5$ part is given by\footnote{Again note that when $\omega = 0$ this part of the geometry becomes that of $S^5$ in standard coordinates after the replacements : $\xi_1 \rightarrow \xi_1 - t/l, \xi_2 \rightarrow \xi_2 - t/l, \xi_3 \rightarrow \xi_3 - t/l$.}
   \begin{eqnarray}
       e^5 &=& l d\alpha ,\nonumber\\
       e^6&=& l\cos\alpha d\beta, \nonumber\\
       e^7 &=& l\cos\alpha\sin\alpha\left[ d\xi_1 -\sin^2 \!\! \beta \,\, d\xi_2 -\cos^2\! \beta \, d\xi_3\right],\nonumber\\
       e^8 &=& l\cos\alpha\sin\beta\cos\beta\left[d\xi_2 - d\xi_3\right],\nonumber\\
       e^9 &=& -\frac{2}{\sqrt{3}}A -l\sin^2\!\! \alpha \,\, d\xi_1 - l\cos^2 \!\! \alpha(\sin^2\!\! \beta \,\, d\xi_2 + \cos^2\!\! \beta \,\, d\xi_3).\label{gr-s5}
   \end{eqnarray}
where the KK gauge potential $A$ is given by \begin{equation}
      A = \frac{\sqrt{3}}{2}\left( (1-\frac{\omega^2}{r^2})dt + \frac{\omega^4}{4 l r^2}\sigma_3^L\right).
  \end{equation}
    Here the ranges of coordinates are : $ 0\le \alpha, \beta\le \pi/2,$ $ 0\le \xi_1, \xi_2, \xi_3\le 2\pi$. The self-dual 5-form field strength for this background is given by
    \begin{eqnarray}
        F^{(5)} &= &- \frac{4}{l} ( e^0 \wedge e^1\wedge e^2 \wedge e^3 \wedge e^4 +e^5 \wedge e^6\wedge e^7 \wedge e^8 \wedge e^9) \nonumber\\
        &+&\left[-\frac{\omega^4}{l r^4} (e^0 \wedge e^1\wedge e^4 - e^2 \wedge e^3\wedge e^9) + \frac{\omega^2}{l r^4}(2 r^2 + \omega^2)(e^0 \wedge e^2\wedge e^3 -e^1 \wedge e^4\wedge e^9)\right.\nonumber\\
        &&~+ \left.\frac{2\omega^2\sqrt{l^2 +2\omega^2 +r^2}}{ l r^3}(e^0 \wedge e^1\wedge e^9 + e^2 \wedge e^3\wedge e^4)\right]\wedge (e^5 \wedge e^7+e^6\wedge e^8). \label{gr-fiveform}
    \end{eqnarray}
The two sets of vielbeins (\ref{gr-ads5}, \ref{gr-s5}) along with the 5-form (\ref{gr-fiveform}) represent the 10d supergravity background that preserves two supersymmetries, with the Killing spinor is given by \cite{Gauntlett:2004cm} (see also \cite{Aharony:2021zkr})
 \begin{eqnarray}
        \epsilon = \sqrt{1-\frac{\omega^2}{r^2}} ~ \exp(-\frac{i}{2} (\xi_1 + \xi_2 + \xi_3)) \epsilon_0 .
    \end{eqnarray} 
Here $\epsilon_0$ is a constant 10d Majorana-Weyl spinor constrained to satisfy the following projections \cite{Gauntlett:2004cm} (with $\eta=1$ there)
    \begin{eqnarray}
        \Gamma^{14} \epsilon_0 = i \epsilon_0, \,\,\,\,\,\,\,\,\,\,\,\,\,\, \Gamma^{23}\epsilon_0 = \Gamma^{57}\epsilon_0 = \Gamma^{68} \epsilon_0 = -i \epsilon_0, \,\,\,\,\,\,\,\,\,\,\, \Gamma^{09}\epsilon_0 = \epsilon_0. \label{projections}
    \end{eqnarray} 
Our aim is to find probe D3-branes in this background that preserve both the supersymmetries of the black hole. We will use the $\kappa$-projection conditions to achieve this, which for the purely geometric embeddings (that is, in the absence of the world-volume gauge field) reads
\begin{eqnarray}
    \gamma_{\tau\sigma_1\sigma_2\sigma_3}\epsilon=\pm i \sqrt{-h}~ \epsilon \label{kappa}
\end{eqnarray}
where $h$ is the determinant of the induced metric $h_{ij}=\mathfrak{e}^a_i \mathfrak{e}^b_j \eta_{ab}$ on the D3-brane, and $\gamma_{\tau\sigma_1\sigma_2\sigma_3} = \mathfrak{e}^a_{\tau}\, \mathfrak{e}^b_{\sigma_1}\mathfrak{e}^c_{\sigma_2}\mathfrak{e}^d_{\sigma_3}\Gamma_{abcd}$ (the $\pm$ signs indicate whether we are working with a D3-brane or an anti-D3-brane), which in turn is written in terms of the pull-back of all ten one-forms in (\ref{gr-ads5}) and (\ref{gr-s5}) onto the D3 world-volume:
\begin{eqnarray}
        \mathfrak{e}^a_i=e^{a}_{\mu}\partial_iX^{\mu}. \label{pull}
    \end{eqnarray}
Here the world-volume coordinates are $(\sigma_0=\tau, \sigma_1, \sigma_2, \sigma_3) $ represented by the index $i$, and ten coordinates of the background are represented by $X^{\mu}$, where $\mu =0,\cdots,9$. Then the world-volume gamma matrices are \begin{eqnarray}
        \gamma_i=\mathfrak{e}^a_i \Gamma_a.
    \end{eqnarray}
Following \cite{Ashok:2008fa},  we define the following 1-forms that will help us to write down all equations in more compact form:
  \begin{eqnarray}
      \mathbi{E}^1 = \mathfrak{e}^1 + i \mathfrak{e}^4, \,\,\,\,\,\,\,\,\,\,\,\,\,\,\,\, \mathbi{E}^2=\mathfrak{e}^2 - i\mathfrak{e}^3, \nonumber\\
      \mathbi{E}^5=\mathfrak{e}^5 - i \mathfrak{e}^7, \,\,\,\,\,\,\,\,\,\,\,\,\,\,\,\,\,  \mathbi{E}^6=\mathfrak{e}^6 - i\mathfrak{e}^8, \nonumber\\
      \mathbi{E}^0=\mathfrak{e}^0+\mathfrak{e}^9, \,\,\,\,\,\,\,\,\,\,\,\,\,\,\,\,\,\,  \mathbi{E}^{\bar 0}=\mathfrak{e}^0-\mathfrak{e}^9 .
\end{eqnarray}
Along with these we also define two special 2-forms:
\begin{eqnarray}
    \Tilde{\omega}_2= \mathfrak{e}^{23} - \mathfrak{e}^{14}, \,\,\,\,\,\,\,\,\,\,\,\,\,\,\,\,\,\,\,\,\,\,\,\,\,\,\,\,\,\, {\omega}_2 = \mathfrak{e}^{57}+\mathfrak{e}^{68}.
\end{eqnarray}
Now one can use the projections in (\ref{projections}) to simplify the $\kappa$-projection condition. This will provide some differential constraints on the embedding coordinates $X^{\mu} (\sigma_i)$. The RHS of (\ref{kappa}) does not contain any gamma matrices, so the terms containing at least one gamma matrix on the LHS should vanish. When we simplify the LHS of (\ref{kappa}) using (\ref{projections}), it will give three types of terms: (i) terms with no gamma matrices, (ii) terms with product of two gamma matrices and (iii) terms with product of four gamma matrices acting on $\epsilon_0$.  To satisfy the $\kappa$-projection condition the coefficients of each of the terms belonging to classes (ii) and (iii) have to vanish.\footnote{To see why one gets many more conditions than the number of independent parameters in $\epsilon_0$, which is just two in this case, note that there is a complete set of 16 orthogonal/commuting projection operators in the problem, namely $$P_{\eta_1\eta_2 \eta_5 \eta_6} := \tfrac{1-i \eta_1 \Gamma^{14}}{2}\tfrac{1+i \eta_2 \Gamma^{23}}{2}\tfrac{1+i \eta_5 \Gamma^{57}}{2}\tfrac{1+i \eta_6 \Gamma^{68}}{2}$$ for $\eta_i = \pm 1$. The spinor $\epsilon_0$ belongs to the subspace corresponding to the projector $P_{++++}$ and annihilated by any of the other 15. One can now hit the $\kappa$-projection condition with each of these projectors and demand that the coefficient of non-vanishing spinor components have to vanish. One can see that each of the terms belonging to classes (ii) and (iii) are left invariant by one or the other projector in this list with at least one $\eta_i$ negative. This procedure clearly is expected to give rise to a total of 16 (complex) conditions.}
The terms in class (iii) give 
\begin{eqnarray}
    \mathbi{E}^{1256} \, \Gamma_{4256} \epsilon &=&0,\cr
    \mathbi{E}^{0125} \,\Gamma_{0125} \epsilon&=&0, ~~~ \mathbi{E}^{0126} \, \Gamma_{0126} \epsilon=0 \cr
   \mathbi{E}^{0256} \, \Gamma_{0256} \epsilon &=& 0, ~~~ \mathbi{E}^{0156} \, \Gamma_{0156} \epsilon=0.
\end{eqnarray}
which imply that we have to impose the following five conditions
\begin{eqnarray}
    \mathbi{E}^{1256} = 0 ,\,\;\;\;\;\;\;\;\;\;\;\;  \mathbi{E}^{0ABC} =0 \label{fourgamma}
\end{eqnarray}
 for $A, B, C = 1, 2, 5, 6$. The terms in class (ii) give  
 \begin{eqnarray}
      (\mathfrak{e}^{14}-\mathfrak{e}^{23}-\mathfrak{e}^{57}-\mathfrak{e}^{68})\wedge \mathbi{E}^{0A}\Gamma_{0A}\epsilon &=&0, \cr
       (-\mathfrak{e}^{09}+i\mathfrak{e}^{14}-i\mathfrak{e}^{23}-i\mathfrak{e}^{57}-i\mathfrak{e}^{68})\wedge\mathbi{E}^{AB}\Gamma_{AB}\epsilon &=& 0,
 \end{eqnarray}
 where $A, B, C = 1, 2, 5, 6$, and repeated indices are not summed over. Thus we arrive at the following ten conditions
\begin{eqnarray}
    \left(\mathfrak{e}^{09} + i (\Tilde{{\omega}_2}+{\omega}_2)\right) \wedge \mathbi{E}^{AB} = 0 ~~~~ {\rm for} ~~~A, B=0,1,2,5,6. \label{twogamma}
\end{eqnarray} 
The remain terms are independent of gamma matrices and these are \begin{eqnarray}
    \mathfrak{e}^{09}\wedge (-i\mathfrak{e}^{14}+i\mathfrak{e}^{23}+i\mathfrak{e}^{57}+i\mathfrak{e}^{68})\epsilon , \;\;\;\;\; (\mathfrak{e}^{1423} +\mathfrak{e}^{1457}+\mathfrak{e}^{1468}-\mathfrak{e}^{2357}-\mathfrak{e}^{2368}-\mathfrak{e}^{5768})\epsilon\nonumber
\end{eqnarray}
Using (\ref{fourgamma}, \ref{twogamma}), the $\kappa$-projection condition reduces to 
\begin{eqnarray}
     \mathfrak{e}^{09} \wedge (\Tilde{{\omega}_2}+{\omega}_2) +\frac{i}{2}(\Tilde{{\omega}_2}+{\omega}_2)\wedge(\Tilde{{\omega}_2}+{\omega}_2) = \pm \sqrt{-h}. \label{gammaindependent}
 \end{eqnarray}
For simplifying the RHS of (\ref{gammaindependent}) using the BPS conditions one can show (following manipulations similar to those in \cite{Ashok:2008fa}) that
\begin{eqnarray}
     h = -\left((\Tilde{{\omega}_2}+{\omega}_2)\wedge\mathfrak{e}^{09}\right)^2 +\frac{1}{4}\left((\Tilde{{\omega}_2}+{\omega}_2)\wedge(\Tilde{{\omega}_2}+{\omega}_2)\right)^2.
\end{eqnarray}
Finally, to solve (\ref{gammaindependent}) we restrict to the time-like D3-branes where we further impose the condition,\begin{eqnarray}
\label{timelike-cond}
    (\Tilde{{\omega}_2}+{\omega}_2)\wedge (\Tilde{{\omega}_2}+{\omega}_2) = 0.
\end{eqnarray}
Then the $\kappa$-projection condition will be satisfied for a D3-brane (anti-brane) for positive (negative) sign of $\mathfrak{e}^{09} \wedge (\tilde{{\omega}_2}+{\omega}_2)$.  Thus we arrive at the full set of conditions (\ref{fourgamma}), (\ref{twogamma}) and (\ref{timelike-cond}) for the embedding coordinates $X^\mu(\sigma_i)$ of a D3-brane to preserve both the supersymmetries of the black hole. 

Remarkably, the BPS equations we have just obtained have the same form as those of \cite{Ashok:2008fa} (after appropriate relabelings). As we will see, these equations can be solved in their generality to obtain all Mikhailov giant and wobbling dual-giant type supersymmetric embeddings of probe D3-branes in the GR black hole as well. In case of giants and dual-giant these conditions can be simplified further. 

For the giants using the fact that their world-volume extends along one dimension in the (asymptotically) AdS directions, any 4-form with more than one index from these directions when pulled back onto the world-volume vanishes. Using this one arrives at the following conditions
\begin{equation}
    \mathbi{E}^0 \wedge \mathbi{E}^{56}\wedge \mathbi{E}^1 =0, \;\;\;\;\;\;\;\;\;\; \mathbi{E}^0 \wedge \mathbi{E}^{56}\wedge \mathbi{E}^2 =0, \label{gc1}
\end{equation}
\begin{eqnarray}
    \left[\begin{array}{c}
  \mathbi{E}^0\\ \mathbi{E}^5 \\ \mathbi{E}^6
\end{array}
\right] \wedge \left[ \begin{array}{c}
  \mathbi{E}^0\\ \mathbi{E}^1 \\ \mathbi{E}^2
\end{array}
\right] \wedge {\omega}_2 = 0, \label{gc3}
\end{eqnarray}
\begin{equation}
     \mathfrak{e}^{09} \wedge \mathbi{E}^{56} =0, ~~ {\omega}_2\wedge{\omega}_2 = 0 \label{gc4}
 \end{equation}
Similarly, for dual-giants (that extend along just one direction in $S^5$) the BPS conditions become
\begin{eqnarray}
    \mathbi{E}^0 \wedge \mathbi{E}^{12}\wedge \mathbi{E}^5 &=& 0, ~~~
    \mathbi{E}^0 \wedge \mathbi{E}^{12}\wedge \mathbi{E}^6 = 0, \nonumber\\
    \left[\begin{array}{c}
  \mathbi{E}^0\\ \mathbi{E}^5 \\ \mathbi{E}^6
\end{array}
\right] \wedge \left[ \begin{array}{c}
  \mathbi{E}^0\\ \mathbi{E}^1 \\ \mathbi{E}^2
\end{array}
\right] \wedge \Tilde{{\omega}_2} &=& 0, \nonumber\\
\mathfrak{e}^{09} \wedge \mathbi{E}^{12} &=& 0, ~~~
\Tilde{ {\omega}_2} \wedge \Tilde{{\omega}_2 } = 0. \label{dgg}
\end{eqnarray}
We are now ready to solve these equations. 
\section{The dual-giant solutions}
\label{dualgiantsection}
To find the dual-giants we impose the  following static gauge
\begin{eqnarray}
    t = \tau, \;\;\; \theta =\sigma_1,\;\;\; \phi = \sigma_2,\;\;\;\; \psi = \sigma_3, \end{eqnarray}
and solve the conditions (\ref{dgg}) to constrain all transverse $X^{\mu}$ as functions of world-volume coordinates. The fact that a dual-giant shares at most one direction in $S^5$ with its world-volume implies that any 2-form of $S^5$ will have to pull-back to zero. This necessitates that all five transverse coordinates in $S^5$ be functionals of a single real function $f$ of the world-volume coordinates:
\begin{eqnarray}
    \alpha=\alpha(f( \tau, \sigma_i)),\;\;\;\;\; \beta = \beta(f( \tau, \sigma_i)),\;\;\;\;\;\; \xi_i=\xi_i(f( \tau, \sigma_i))
\end{eqnarray}
Using (\ref{pull}) we can write down the pull-back 1-forms onto the dual-giant as:
\begin{eqnarray}
    \mathfrak{e}^0 &=& \left(1-\frac{\omega^2}{r^2} \right) dt + \frac{\omega^4-2 r^4 -2 r^2 \omega^2 }{4 l r^2} (d\phi +  \cos\theta \, d\psi),\nonumber\\
    \mathfrak{e}^1 &=& \frac{r^2 l}{(r^2 - \omega^2)\sqrt{l^2 + r^2 +2\omega^2}}\left(\dot{r} dt + r_{\theta} d\theta + r_{\phi} d\phi +r_{\psi} d\psi \right),\nonumber\\
    \mathfrak{e}^2 &=& \frac{r}{2}\left( \sin\phi d\theta -\cos\phi\sin\theta \, d\psi\right), \nonumber\\
    \mathfrak{e}^3 &=& \frac{r}{2}\left(\cos\phi \, d\theta +\sin\theta\sin\phi \, d\psi\right),\nonumber\\
    \mathfrak{e}^4 &=& \frac{r}{2 l}\sqrt{l^2 +r^2 +2\omega^2}\left( d\phi + \cos\theta \, d\psi\right),
     \end{eqnarray}
   \begin{eqnarray} \mathfrak{e}^5 &=& l \, \alpha^\prime \, df,\nonumber\\
    \mathfrak{e}^6 &=& l \, \cos\alpha \, \beta^\prime \, df,\nonumber\\
    \mathfrak{e}^7 &=& l \, \cos\alpha\sin\alpha\left(\xi_1^\prime - \sin^2\beta \,\xi_2^\prime -\cos^2\beta \, \xi_3^\prime\right) \, df,\nonumber\\
    \mathfrak{e}^8 &=& l \, \cos\alpha\cos\beta\sin\beta \, (\xi_2^\prime - \xi_3^\prime) \, df,\\
    \mathfrak{e}^9 &=& -\left( 1 -\frac{\omega^2}{r^2} \right) \, dt -\frac{\omega^4}{4 l r^2} (d\phi +\cos\theta \, d\psi) \cr
    && -l \, (\sin^2\alpha \, \xi_1^\prime + \cos^2\alpha\sin^2\beta \, \xi_2^\prime + \cos^2\alpha\cos^2\beta \, \xi_3^\prime) \, df  \nonumber
\end{eqnarray}
where we $\alpha^\prime = \frac{\delta \alpha(f)}{\delta f}$ and so on.
 We start by imposing the last condition of (\ref{dgg}), which gives rise to the following (whenever $r\ne 0$ and $r\ne \omega$) , \begin{eqnarray}
     \mathfrak{e}^{1234} =0 &\Longrightarrow& \frac{r^5 \dot{r}\sin\theta}{8(r^2 - \omega^2)} dt \wedge d\theta \wedge d\phi \wedge d\psi =0 \cr \cr
   &\Longrightarrow& \dot{r} = 0 .
 \end{eqnarray}
The imaginary part of the condition $ \mathbi{E}^{01} \wedge \Tilde{\omega} = (\mathfrak{e}^{0123}+\mathfrak{e}^{9123}) + i (\mathfrak{e}^{0423}+\mathfrak{e}^{9423}) = 0$  gives  
\begin{eqnarray}
   \frac{r^3\sin\theta}{8}\sqrt{l^2+r^2+2\omega^2} \left(\sin^2\alpha\, \xi_1^\prime + \cos^2\alpha\sin^2\beta \, \xi_2^\prime + \cos^2\alpha\cos^2\beta \, \xi_3^\prime \right) \dot f =0.
\end{eqnarray}
This can be solved for generic values of $r, \alpha, \beta$ by 
\begin{eqnarray}
\xi_1^\prime= \xi_2^\prime = \xi_3^\prime = 0~~~ {\rm or}~~~ \dot f=0
\end{eqnarray}
We will use $\dot f =0$, {\it i.e.}, $f(\tau, \sigma_i) = f(\sigma_i)$ -- as this choice will lead to more general solutions, and will end up including the former. Then it easily follows that $(\mathfrak{e}^0+\mathfrak{e}^9) \wedge \mathfrak{e}^{123} =0$ as well. In fact all 4-form conditions which involve $\mathbi{E}^0$ will pull-back to zero as none of them can have $d\tau$ as $\dot r = \dot f =0$. Similarly, the equations $ \mathbi{E}^{AB} \wedge \Tilde{{\omega}_2} = 0$ for $A,B = 1,2,5,6$ are trivially satisfied. The remaining conditions are  $ \mathfrak{e}^{09}\wedge \mathbi{E}^{AB} = 0$, which are trivially satisfied for $A,B=5,6$. The case $A,B=1,2$ will be treated separately later. Rest of the conditions (for $A \in \{1,2\}$ and $B \in \{5,6\}$) lead to the following constraints
\begin{eqnarray}
     \xi_1^\prime (f)= \xi_2^\prime(f) = \xi_3^\prime(f) \equiv \xi'(f) \;\;\;\;\;\; {\rm and} \;\;\;\;\;\; \alpha'(f) = \beta'(f) = 0.
\end{eqnarray}
 The constraint on $r$ comes from $ \mathfrak{e}^{09}\wedge \mathbi{E}^{12} =(\mathfrak{e}^{0912} +\mathfrak{e}^{0943})+i(\mathfrak{e}^{0913} -\mathfrak{e}^{0942} )=0 $. This will turn into the following equation
\begin{eqnarray}
 & & \xi'(r^2-\omega^2)(l^2+r^2+2\omega^2)\left[(f_{\psi} - f_{\phi} \cos\theta) \cos\phi - f_{\theta}\sin\theta\sin\phi\right] \nonumber\\
  &-& r (r^2 + 2 f_{\phi}l^2 \xi' +\omega^2)\sin\phi\frac{\partial r}{\partial \psi} + r (r^2 + 2 f_{\phi}l^2 \xi' +\omega^2)\sin\theta\cos\phi\frac{\partial r}{\partial \theta} \nonumber\\
  &-& r[2f_{\theta}l^2\xi'\cos\phi\sin\theta + (2f_{\psi}l^2 \xi' +(r^2+\omega^2)\cos\theta)\sin\phi]\frac{\partial r}{\partial \phi}\cr
 & +i & (\xi'(r^2-\omega^2)(l^2+r^2+2\omega^2)\left[(f_{\psi} - f_{\phi} \cos\theta) \sin\phi + f_{\theta}\sin\theta\cos\phi\right] \nonumber\\
  &+i& \left(r (r^2 + 2 f_{\phi}l^2 \xi' +\omega^2)\cos\phi\frac{\partial r}{\partial \psi} - r (r^2 + 2 f_{\phi}l^2 \xi' +\omega^2)\sin\theta\sin\phi\frac{\partial r}{\partial \theta}\right) \nonumber\\
  &+i& r[2f_{\theta}l^2\xi'\sin\phi\sin\theta - (2f_{\psi}l^2 \xi' +(r^2+\omega^2)\cos\theta)\cos\phi]\frac{\partial r}{\partial \phi} = 0.
    \end{eqnarray}
Now changing the variable 
\begin{eqnarray}
    r^2 = \omega^2 + (l^2+3\omega^2)\;\sinh^2\rho \;,\label{radial defination}
\end{eqnarray} 
one can rewire it as
\begin{eqnarray}
&&(l^2+3\omega^2)\sinh\rho \cosh\rho \left( \frac{\partial\xi}{\partial\psi} - \frac{\partial\xi}{\partial\phi}\cos\theta + i \sin\theta \frac{\partial\xi}{\partial\theta}\right) \cr
&+& i \left((l^2+3\omega^2) \sinh^2\rho+2\omega^2 +2 l^2 \frac{\partial\xi}{\partial\phi}\right)\frac{\partial\rho}{\partial\psi} \cr
   &-&i  \left([(l^2+3\omega^2) \sinh^2\rho+2\omega^2]\cos\theta+2l^2\left(\frac{\partial\xi}{\partial\psi} +i \frac{\partial\xi}{\partial\theta}\sin\theta\right)\right) \frac{\partial\rho}{\partial\phi} \cr
   &-&\left[(l^2+3\omega^2) \sinh^2\rho+2\omega^2+2l^2\frac{\partial\xi}{\partial\phi}\right]\sin\theta\frac{\partial\rho}{\partial\theta} = 0.\label{general eq1}
\end{eqnarray}
To find general solutions we switch to the embedding function language and assume that the world-volume is given by simultaneous zeros of two real functions $\rm{f}$ and $\rm{g}$ of the five coordinates $\rho, \theta, \phi, \psi, \xi$, as
 \begin{eqnarray}
    \rm f(\rho, \theta, \phi, \psi, \xi) = 0~~~~~~~~{\rm and} ~~~~~~~~ \rm g(\rho, \theta, \phi, \psi, \xi) = 0 \, .
\end{eqnarray} 
Taking the differential of these functions one should have the following \begin{eqnarray}
    && \rm f_{\rho} d\rho + \rm f_{\theta} d\theta + \rm f_{\phi} d\phi + \rm f_{\psi} d\psi + \rm f_{\xi} d{\xi} = 0, \cr 
  &&  \rm g_{\rho} d\rho + \rm g_{\theta} d\theta + \rm g_{\phi} d\phi + \rm g_{\psi} d\psi + \rm g_{\xi} d{\xi} = 0,
\end{eqnarray}
where $\rm f_\rho= \frac{\partial \rm f}{\partial\rho}$ etc. We choose to solve these to write ${\rm d\rho}$ and ${\rm d\xi}$ in terms of $ {\rm d\theta, d\phi, d\psi}$, which results in the following 
\begin{eqnarray}
    \frac{\partial \rho}{\partial \theta} = \frac{\rm f_{\xi}g_{\theta}-f_{\theta}\rm g_{\xi}}{\rm f_{\rho}\rm g_{\xi}- \rm f_{\xi}\rm g_{\rho}},~~~~~~     \frac{\partial \rho}{\partial \phi} = \frac{\rm f_{\xi}\rm g_{\phi}-\rm f_{\phi}\rm g_{\xi}}{\rm f_{\rho}\rm g_{\xi}- \rm f_{\xi}\rm g_{\rho}},~~~~~~    \frac{\partial \rho}{\partial \psi} = \frac{\rm f_{\xi}\rm g_{\psi}-\rm f_{\psi}\rm g_{\xi}}{\rm f_{\rho}\rm g_{\xi}- \rm f_{\xi}\rm g_{\rho}},~~~~~~\cr
        \frac{\partial \xi}{\partial \theta} = \frac{\rm f_{\rho}\rm g_{\theta}-\rm f_{\theta}\rm g_{\rho}}{-\rm f_{\rho}\rm g_{\xi}+ \rm f_{\xi}\rm g_{\rho}},~~~~~~ \frac{\partial \xi}{\partial \phi} = \frac{\rm f_{\rho}\rm g_{\phi}-\rm f_{\phi}\rm g_{\rho}}{-\rm f_{\rho}\rm g_{\xi}+ \rm f_{\xi}\rm g_{\rho}},~~~~~~ \frac{\partial \xi}{\partial \psi} = \frac{\rm f_{\rho}\rm g_{\psi}-\rm f_{\psi}\rm g_{\rho}}{-\rm f_{\rho}\rm g_{\xi}+ \rm f_{\xi}\rm g_{\rho}}.\label{relations1}
\end{eqnarray}
Substituting these in (\ref{general eq1}) we arrive at
 \begin{eqnarray}
   && (l^2+3\omega^2) \sinh\rho\cosh\rho((\rm f_{\psi}\rm g_{\rho}-\rm f_{\rho}\rm g_{\psi}) +(\rm f_{\rho}\rm g_{\phi}- \rm f_{\phi}\rm g_{\rho})\cos\theta -i\;\sin\theta(\rm f_{\rho}\rm g_{\theta}-\rm g_{\rho}\rm f_{\theta})) \cr &-& i(2\omega^2+ (l^2+3\omega^2)\sinh^2\rho)(( \rm f_{\psi}\rm g_{\xi}-\rm f_{\xi}\rm g_{\psi})+\cos\theta(\rm f_{\xi}\rm g_{\phi} -\rm f_{\phi}\rm g_{\xi})-i\;\sin\theta (\rm f_{\xi}\rm g_{\theta}-\rm f_{\theta}\rm g_{\xi}))\cr &+& 2i \;l^2 ((\rm f_{\psi}\rm g_{\phi}-\rm f_{\phi}\rm g_{\psi})-i\;\sin\theta (\rm f_{\phi}\rm g_{\theta}-\rm f_{\theta}\rm g_{\phi})) = 0.
    \end{eqnarray}
This can be recast in very simple form as:
\begin{eqnarray}
        X(\rm f) Y(\rm g) - X(\rm g) Y(\rm f) = 0\label{vector field equation}
\end{eqnarray}
    where $X$ and $Y$ are the following differential operators 
    \begin{eqnarray}
    \label{XYdefs}
      X &=&  (l^2+3\omega^2) \sinh\rho\cosh\rho\; \frac{\partial}{\partial \rho} -i\;(2\omega^2 +(l^2+3\omega^2)\sinh^2\rho ) \frac{\partial}{\partial\xi} + 2i\;l^2 \frac{\partial}{\partial \phi}\cr
      Y &=& -i\;\sin\theta \frac{\partial}{\partial\theta} +\cos\theta\frac{\partial}{\partial\phi} -\frac{\partial}{\partial\psi}.
    \end{eqnarray}
    Note that in arriving at this equation we have assumed \begin{eqnarray}
    \rm f_{\rho}\rm g_{\xi}- \rm f_{\xi}\rm g_{\rho} \ne 0 \, .\label{important condition}
\end{eqnarray}
  We can also assumed that the world-volume is given by the zeros of a single complex function $F= \rm f +i \, \rm g$, and write (\ref{vector field equation}) as  
  \begin{eqnarray}
         X(F) Y(\bar F) - X(\bar F) Y(F) = 0.\label{complex vector field equation}
   \end{eqnarray}
In this case one should remember that the condition (\ref{important condition}) $F_\rho \bar F_\xi - F_\xi \bar F_\rho \ne 0$ dictates that $ F $ must depend on both $\rho$ and $\xi$, in a non-trivial way. 

So far we have analysed our BPS conditions for a dual-giant in a particular gauge. To obtain the full set of solutions one should like to do a gauge independent analysis. Let us discuss the derivation of this kind of equation without making a choice of a gauge.
\subsection*{Analysis without gauge choice} \label{dual giant solutions}
Suppose we try to find the general solution for BPS D3-branes in our black hole background by defining  general embedding functions. In the general case one needs three independent complex constraints which specify the world-volume for the D3-brane. Any one of these constraints can taken to be
\begin{eqnarray}
     F ( t,\; r, \;\theta, \;\phi,\; \psi, \;\alpha, \;\beta,\; \xi_1,\; \xi_2,\; \xi_3) = 0 \, .
 \end{eqnarray}
Then taking the differential of this condition, the pull-backs of the spacetime 1-forms $e^a$ onto the world-volume have to satisfy
 \begin{eqnarray}
     &&\frac{1}{2}\left[  -\frac{2}{l}(F_{\xi_1} +F_{\xi_2}+F_{\xi_3}) + \frac{F_t}{(1-\frac{\omega^2}{r^2})}\right] \mathbi{E}^0 +\frac{1}{2} \frac{F_t}{(1-\frac{\omega^2}{r^2})} \mathbi{E}^{\bar0} \nonumber\\
    && +\frac{1}{2} \left[F_r \sqrt{\frac{r^2 + l^2 +2\omega^2}{l^2}}(1-\frac{\omega^2}{r^2}) \right. \nonumber\\ &&\left.- i \frac{2(\omega^4 - r^4) (F_{\xi_1} +F_{\xi_2}+F_{\xi_3})+ 2 F_t l (r^4 +r^2\omega^2 -\frac{\omega^4}{2}) +4 F_{\phi} l^2 (r^2 - \omega^2)}{2 r l \sqrt{l^2+ r^2+ 2\omega^2} (r^2 - \omega^2)}\right]\mathbi{E}^1 \nonumber\\
    && + \frac{1}{2} \left[F_r \sqrt{\frac{r^2 + l^2 +2\omega^2}{l^2}}(1-\frac{\omega^2}{r^2}) \right. \nonumber\\&& \left. +i \frac{2(\omega^4 - r^4) (F_{\xi_1} +F_{\xi_2}+F_{\xi_3})+ 2 F_t l (r^4 +r^2\omega^2 -\frac{\omega^4}{2}) +4 F_{\phi} l^2 (r^2 - \omega^2)}{2 r l \sqrt{l^2+ r^2+ 2\omega^2} (r^2 - \omega^2)}\right]{\mathbi{E}^{\bar 1}}  \nonumber\\
   &&  + \frac{1}{r}e^{-i \phi}\left[ (F_{\phi}\;\cot\theta - F_{\psi} \;\csc\theta +i\; F_{\theta} ) \right]\mathbi{E}^2 + \frac{1}{r}\;e^{i\phi}\left[ (F_{\phi}\;\cot\theta - F_{\psi} \;\csc\theta -i\; F_{\theta} ) \right]\mathbi{E}^{\bar 2 } \nonumber\\
  &&   + \frac{1}{2}\left[ \frac{F_{\alpha}}{l} +i \frac{1}{l}( F_{\xi_1} \cot\alpha - F_{\xi_2}\tan\alpha - F_{\xi_3}\tan\alpha) \right]\mathbi{E}^5 \nonumber\\ &&+ \frac{1}{2}\left[ \frac{F_{\alpha}}{l} -i \frac{1}{l}( F_{\xi_1} \cot\alpha - F_{\xi_2}\tan\alpha - F_{\xi_3}\tan\alpha) \right]\mathbi{E}^{\bar 5}  \nonumber\\
 &&  + \frac{1}{2l}\sec\alpha  \left[ F_{\beta} +i  \, (F_{\xi_2}\cot\beta - F_{\xi_3}\tan\beta) \right]\mathbi{E}^6 \nonumber\\ && + \frac{1}{2l}\sec\alpha  \left[ F_{\beta} - i  \, (F_{\xi_2}\cot\beta - F_{\xi_3}\tan\beta) \right]\mathbi{E}^{\bar 6} = 0. \label{general constraint}
     \end{eqnarray}
Here we have written $\mathbi{E}^{\bar a} = (\mathbi{E}^a)^\star$ for $a=1,2,5,6$. Since this is a complex condition, its conjugate should also vanish. This will provide another equation as (\ref{general constraint}) where $F$ is replaced by $\bar F$.  To get 4d world-volume  we need to  consider two more functions and their consequent 1-form constraints. 
     
For the rest of this section we restrict to the dual-giant case, and simply take the required two 1-form constraints to be \begin{eqnarray}\mathbi{E}^5 = \mathbi{E}^6 = 0,\label{two constraint}\end{eqnarray}
as these were true  in the static gauge analysis of the previous subsection, which imply \begin{eqnarray}
          {\bf d\alpha} = {\bf d\beta} = 0, \;\;\;\;\;\;\;\;\;\;\;\;\;\;\;\;\;  {\bf d\xi_1 = d\xi_2 = d\xi_3}.
      \end{eqnarray}
      These differential conditions require \begin{eqnarray}
          \alpha = \alpha_0,\;\;\;\;\;\; \beta = \beta_0, \;\;\;\;\;\; \xi_1 - \xi_2 = \xi_{12}^{(0)}, \;\;\;\;\;\; \xi_1 - \xi_3 = \xi^{(0)}_{13}. \label{four constants}
      \end{eqnarray}
Substituting these four constraints into the remaining one embedding function becomes $F= F(t,\;r,\;\theta,\;\phi,\,\psi,\;\xi)$ where  $\xi =(\xi_1 + \xi_2 + \xi_3 )/3$. 
Let us further define the following vector fields 
\begin{eqnarray}
\label{diffopsX}
     X_{\bar0} &=& \frac{1}{(1-\frac{\omega^2}{r^2})}\frac{\partial}{\partial t}\nonumber\\
    X_0 &=& -\frac{2}{l}\frac{\partial}{\partial \xi} + \frac{1}{(1-\frac{\omega^2}{r^2})}\frac{\partial}{\partial t}\nonumber\\
    X_1 &=&  \sqrt{\frac{r^2 + l^2 +2\omega^2}{l^2}}(1-\frac{\omega^2}{r^2})\frac{\partial}{\partial r} \cr
    && ~~~~ - i \frac{2(\omega^4 - r^4) \frac{\partial}{\partial\xi}+ 2  l (r^4 +r^2\omega^2 -\frac{\omega^4}{2})\frac{\partial}{\partial t} +4 l^2 (r^2 - \omega^2)\frac{\partial}{\partial \phi}}{2 r l \sqrt{l^2+ r^2+ 2\omega^2} (r^2 - \omega^2)} \nonumber\\
    X_2& =&\;\frac{2}{r}e^{-i \phi}[\cot\theta \frac{\partial}{\partial\phi}-  \;\csc\theta \frac{\partial}{\partial\psi} +i\; \frac{\partial}{\partial\theta}]
\end{eqnarray}
 in terms of which the equation (\ref{general constraint}) for $F$ and $\bar F$ can be written as 
 \begin{eqnarray}
     X_1(F) \mathbi{E}^1+ \bar{X_1}(F)  \bar{\mathbi{E}^1}+ X_2(F) \mathbi{E}^2+ \bar{X_2}(F)  \bar{\mathbi{E}^2}+ X_0(F) \mathbi{E}^0+ X_{\bar0}(F)  \mathbi{E}^{\bar0} &=& 0 \, , \nonumber\\
 X_1(\bar F) \mathbi{E}^1+ \bar{X_1}(\bar F)  \bar{\mathbi{E}^1}+ X_2(\bar F) \mathbi{E}^2+ \bar{X_2}(\bar F)  \bar{\mathbi{E}^2}+ X_0(\bar F) \mathbi{E}^0+ X_{\bar0}(\bar F)  \mathbi{E}^{\bar0} &=& 0.  \,\label{equ52}
 \end{eqnarray}
Now one can solve these equations  for any two 1-forms and substitute them in any of the BPS conditions. Then using the other BPS conditions one can get further differential equations only from the BPS conditions which are not trivially satisfied. 

Following these steps, solving (\ref{equ52}) for $E^0$ and $E^1$ and substituting in the BPS condition $E^{0\bar0 12}= 0$, one can get 
 \begin{eqnarray}
    [ \bar{X_1}(F)\bar{X_2}(\bar F) - \bar{X_1}(\bar F) \bar{X_2}(F) ] E^{\bar0\bar1\bar2 2} =0.\label{conseq1}
 \end{eqnarray}
In a similar way substituting $E^1$ and $E^{\bar1}$ in the BPS condition $E^{1\bar1 2\bar2} = 0$, one can get 
  \begin{eqnarray}
      [X_{\bar0}(\bar F) X_0(F) - X_{\bar0} (F) X_0(\bar F)] E^{0\bar0 1 \bar1} = 0.\label{conseq2}
  \end{eqnarray}
Substituting $E^1$ and $E^{\bar1}$ in the BPS condition $E^{01\bar1 2} = 0$ leads to
\begin{eqnarray}
     [X_{\bar0}(\bar F) \bar X_2(F) - X_{\bar0} (F) \bar X_2(\bar F)] E^{0\bar0 2 \bar2} = 0.\label{conseq3}
\end{eqnarray}
 Another condition can be found from the BPS condition $E^{02\bar21} = 0$,
 \begin{eqnarray}
       [X_{\bar0}(\bar F) \bar X_1(F) - X_{\bar0} (F) \bar X_1(\bar F)] E^{0\bar0 1 \bar1} = 0.\label{conseq4}
 \end{eqnarray}
Since the remaining BPS conditions are just the complex conjugates of the BPS conditions used above, they will only provide the complex conjugates of these equations. Since the equations (\ref{conseq1} - \ref{conseq4}) contain 4-forms that do not necessarily vanish by the BPS  conditions, their coefficients must vanish, resulting in
\begin{eqnarray}
    X_{\bar0}(\bar F) X_0(F) - X_{\bar0} (F) X_0(\bar F)&=&0\, , \label{diffeq1}\\
    X_{\bar0}(\bar F) \bar X_1(F) - X_{\bar0} (F) \bar X_1(\bar F)&=& 0 \, , \label{diffeq2}\\
    X_{\bar0}(\bar F) \bar X_2(F) - X_{\bar0} (F) \bar X_2(\bar F) &=& 0 \, ,\label{diffeq3}\\
    \bar{X_1}(F)\bar{X_2}(\bar F) - \bar{X_1}(\bar F) \bar{X_2}(F)&=& 0 \, ,\label{diffeq4}
\end{eqnarray}
along with
\begin{eqnarray}
    X_0(F) \bar{X_1}(\bar F) -X_0(\bar F)\bar{X_1}(F) \;\ne \; 0,&&\;\;\;\;\;\;\;\;\;  X_0(F) \bar{X_2}(\bar F) -X_0(\bar F)\bar{X_2}(F) \;\ne \; 0, \nonumber\\
    X_2(F) \bar{X_2}(\bar F) - X_2(\bar F) \bar{X_2}(F)\;\ne\;0,&&\;\;\;\;\;\;\;\;X_1(F) \bar{X_1}(\bar F) - X_1(\bar F) \bar{X_1}(F)\;\ne\;0,\nonumber\\
    \bar{X_1}(F) X_2(\bar F) - \bar{X_1}(\bar F) X_2 (F) \;\ne\;0.&&\label{not equal conditions}
\end{eqnarray}
Let us write (\ref{diffeq1}, \ref{diffeq2}) as 
\begin{eqnarray}
\left( \begin{array}{cc} X_0(F) & - X_0(\bar F) \\ \bar X_1(F) & - \bar X_1(\bar F) \end{array} \right) \left( \begin{array}{c} X_{\bar0}(\bar F) \\  X_{\bar0} (F) \end{array}\right) = \left( \begin{array}{c} 0 \\ 0 \end{array}\right) \, 
\end{eqnarray}
which immediately implies 
\begin{eqnarray}
     X_{\bar0}( F) \;=\;0 = X_{\bar0}( \bar F)
\end{eqnarray}
because of the first inequality (\ref{not equal conditions}). Same conclusion can be arrived at by considering (\ref{diffeq1}, \ref{diffeq3}) similarly. As the differential operator $X_{\bar 0}$ is real these two equations are in fact equivalent. 

With these the first three conditions (\ref{diffeq1}-\ref{diffeq3}) are satisfied, and the only remaining condition is the last one $\bar{X_1}(F)\bar{X_2}(\bar F) - \bar{X_1}(\bar F) \bar{X_2}(F)= 0$ (and its complex conjugate), and it is equivalent to  (\ref{complex vector field equation}) arrived at working in the static gauge.

Taking into account the conditions  in (\ref{not equal conditions}), the simplest possibilities to solve equation (\ref{diffeq4}) are
  \begin{eqnarray}
     (i) \; &:&\;\; \bar X_1 (F) = \bar X_2 (F) =0 \, , \cr
     && ~~~~~~~~ {\rm or} \cr
      (ii) \; &:&\;\; \bar X_1 (\bar F) = \bar X_2 (\bar F) =0.\label{4equations}
 \end{eqnarray}
let us find solutions to these equations. Given that the directions $(\phi, \psi, \xi)$ are periodic, one can consider the general ansatz of the form
 \begin{eqnarray}
     F = \sum_{m,n,q} C_{mnq} ( r, \theta) e^{im\phi + i n \psi + i q \xi}.\label{trialf}
 \end{eqnarray}
 where the sum over $(m,n,q)$ are over either ${\mathbb Z}$ or ${\mathbb Z}/2$ depending on the angles involved are periodic with period $2\pi$ or $4\pi$. 

Let us first impose $\bar X_1 (F) = \bar X_2 (F) =0$. After substituting (\ref{trialf}), we obtain
 \begin{eqnarray}
    \partial_r C_{mnq} - \frac{2 m l^2 -(\omega^2 + r^2) q }{r (r^2 + l^2+ 2 \omega^2 ) (1-\frac{\omega^2}{r^2})} C_{mnq} &=& 0\, , \cr && \cr
    \partial_{\theta} C_{mnq} - (m \cot\theta - n \csc\theta) C_{mnq} &=& 0.
\end{eqnarray}
These can be solved completely to obtain
\begin{eqnarray}
    F = \sum_{m,n,q} c_{mnq} \left( \cot\frac{\theta}{2}\; e^{i\psi}\right)^n  &&\left( (r^2 - \omega^2)^{-\frac{\omega^2}{l^2 +3\omega^2}}\;(r^2 +l^2 +2\omega^2)^{-\frac{\omega^2 +l^2}{2(l^2 + 3\omega^2)}}\;\; e^{i\xi}\right)^q \nonumber\\ &\times& \left[\left(\frac{r^2 - \omega^2}{(r^2 + l^2 + 2 \omega^2 )}\right)^{\frac{l^2}{l^2 + 3\omega^2}}(\sin\theta \;e^{i \phi})\right]^m \label{Fconstraint}
\end{eqnarray}
where $c_{mnq}$ are complex numbers. 
This embedding function can be written compactly as $F(\Phi_1, \Phi_2, \Phi_3)$, where
\begin{eqnarray}
    \Phi_1 &=& \cot\frac{\theta}{2}\; e^{i\psi}, \cr
    \Phi_2 &=& (r^2 - \omega^2)^{-\frac{\omega^2}{l^2 +3\omega^2}}\;(r^2 +l^2 +2\omega^2)^{-\frac{\omega^2 +l^2}{2(l^2 + 3\omega^2)}}\;\; e^{i\xi}, \cr
    \Phi_3 &=& \left(\frac{r^2 - \omega^2}{(r^2 + l^2 + 2 \omega^2 )}\right)^{\frac{l^2}{l^2 + 3\omega^2}} \, \sin\theta \;e^{i \phi}.
\end{eqnarray}
To compare this result with dual-giants in $AdS_5 \times S^5$ in \cite{Ashok:2008fa} in the $\omega \rightarrow 0$ limit, we use the following radial coordinate $\rho$ that covers the region outside the horizon: (\ref{radial defination}).
\begin{equation}
r^2-\omega^2 = (l^2+3\omega^2)\, \sinh^2\rho \label{r rho trans}
\end{equation}
and, in terms of which, we define:
\begin{eqnarray}
\frac{1}{\Phi_2} := \Psi_0= \sqrt{l^2+3\omega^2} \, (\sinh\rho)^{\frac{2\omega^2}{l^2+3\omega^2}} \, (\cosh\rho)^{\frac{l^2+\omega^2}{l^2+3\omega^2}} e^{-i\xi}\, , \cr
\sqrt{\frac{\Phi_1\Phi_3}{2\Phi_2^2}} := \Psi_1 = \sqrt{l^2+3\omega^2} \, (\sinh\rho)^{\frac{l^2+2\omega^2}{l^2+3\omega^2}} \, (\cosh\rho)^{\frac{\omega^2}{l^2+3\omega^2}} \, \cos\frac{\theta}{2} \, e^{\frac{i}{2}(\phi+\psi-2\xi)} \, ,\cr
\sqrt{\frac{\Phi_3}{2\Phi_1\Phi_2^2}} := \Psi_2 = \sqrt{l^2+3\omega^2} \, (\sinh\rho)^{\frac{l^2+2\omega^2}{l^2+3\omega^2}} \, (\cosh\rho)^{\frac{\omega^2}{l^2+3\omega^2}} \, \sin\frac{\theta}{2} \, e^{\frac{i}{2}(\phi-\psi-2\xi)}. \label{new complex coordinates}
\end{eqnarray}
In terms of these we can write $F(\Phi_1, \Phi_2, \Phi_3)$ as $G(\Psi_0,\Psi_1, \Psi_2)$ which when $\omega \rightarrow 0$ becomes embedding function for wobbling dual-giants solutions of \cite{Ashok:2008fa} (after some straightforward mapping of coordinates). We propose that just as for the dual-giants in the pure $AdS_5 \times S^5$ background, the holomorphic function $G(\Psi_0,\Psi_1, \Psi_2) = 0$ should be a polynomial in $\Psi_0$ of maximal degree $N$ (the 5-form flux through $S^5$). This reflects the dual-giant version of the stringy exclusion principle as proposed in \cite{Bena:2004qv, Suryanarayana:2004ig}.

Imposing the conditions $\bar X_1 (\bar F) = \bar X_2 (\bar F) =0$ on $F$ in (\ref{trialf}) simply gives the anti-holomorphic $F(\bar\Psi_0,\;\bar\Psi_1,\;\bar\Psi_2)$, which, as a class give completely equivalent solution space to those from $\bar X_1 (F) = \bar X_2 (F) =0$.

To summarise the results so far: $F(\Psi_0, \Psi_1, \Psi_2)=0$ (along with constant $\alpha, \beta, \xi_{ij}$) specify world-volumes of dual-giant type BPS probe D3-branes in the Gutowski-Reall black hole that are generalisations of the corresponding ones in the pure $AdS_5 \times S^5$ background \cite{Ashok:2008fa}.\footnote{We have presented the corresponding result for the 2-parameter generalisation of GR black hole case in Appendix \ref{dgingengr}.} Only a small subclass of these solutions were known earlier \cite{Aharony:2021zkr, Choi:2024xnv}, that correspond to 
\begin{eqnarray}
    r = const., \;\;\;\;\; \xi = const., \label{dualg round}
\end{eqnarray}
which belong to $F(\Psi_0)=0$ subclass in our language, which are precisely the $SU(2)_R$ invariant dual-giants used to form dual dressed black holes in \cite{Choi:2024xnv}. 

Before we move to the BPS giants, some comments are in order. Have we found all possible solutions to our BPS equations or are there are any other classes of solutions to the equation (\ref{diffeq4})? The equation $\bar{X_1}(F)\bar{X_2}(\bar F) - \bar{X_1}(\bar F) \bar{X_2}(F)= 0$ can be rewritten as the singularity condition of the matrix 
\begin{eqnarray}
\left(\begin{array}{cc} \bar X_1(F) & \bar X_1(\bar F) \\ \bar X_2(F)  & \bar X_2(\bar F)  \end{array} \right).
\end{eqnarray}
The above ways (\ref{4equations}) of solving this equation amount to demanding that either of the two columns of this matrix vanishes. However, a much weaker condition would be that the two columns (rows) are linearly dependent, and it is important to check if any viable solutions can be obtained this way that are not already captured by the ones given above. We will postpone further discussion on this issue to Appendix \ref{moresolns}. However, we have been able to show (to a good accuracy, in a perturbative expansion) in Appendix \ref{appendix1}, that the above class (\ref{Fconstraint}) captures all solutions of the BPS dual-giants that can be considered as smooth deformations of the round dual-giant (\ref{dualg round}). 

\section{The giant solutions}
\label{giantsection}

In this section we solve the BPS conditions (\ref{gc1} - \ref{gc4}) for giant graviton type solutions. In particular, we look for D3-branes that expand on the $S^5$ part and are point-like in the $AdS_5$ part of the background. Here we present the analysis without choosing a gauge. 
%
%
A D3-brane that is point-like in the directions $(t,r,\theta, \phi,\psi)$ needs to satisfy 
 \begin{eqnarray}
    \mathbi{E}^1 = \mathbi{E}^2 = 0 \label{giant two constraints}
\end{eqnarray}
 and the equation (\ref{general constraint}). The equations (\ref{giant two constraints}) means the following pull-back conditions \begin{eqnarray}
     {\bf d\phi} = {\bf d\psi} = {\bf d r} = {\bf d\theta} =0 \label{giant1}
 \end{eqnarray}
 which imply $(r,\theta,\phi,\psi)$ are constants. We again postulate that the world-volume is further specified by the zeros of a complex function: $F(t,\;\alpha,\; \beta, \;\xi_1, \;\xi_2,\; \xi_3) =0 $. Following the procedure of the previous section in this case, one can be obtained the following equations.
\begin{eqnarray}
    X_{\bar0}(\bar F) X_0(F) - X_{\bar0} (F) X_0(\bar F)&=&0, \label{gdiffeq1}\\
    X_{\bar0}(\bar F) \bar X_5(F) - X_{\bar0} (F) \bar X_5(\bar F)&=& 0, \label{gdiffeq2}\\
    X_{\bar0}(\bar F) \bar X_6(F) - X_{\bar0} (F) \bar X_6(\bar F) &=& 0, \label{gdiffeq3}\\
    \bar{X_5}(F)\bar{X_6}(\bar F) - \bar{X_5}(\bar F) \bar{X_6}(F)&=& 0.\label{gdiffeq4}
\end{eqnarray} 
Where the $X_5 (F)$ and $X_6 (F)$ are the coefficients of $\mathbi{E}^5$ and $\mathbi{E}^6$ in the equation (\ref{general constraint}) respectively. These are supplemented by non-vanishing conditions similar to (\ref{not equal conditions}) where $\bar{X}_1$ and $\bar X_2$ are replaced by the $\bar X_5$ and $\bar X_6$ respectively. The first three equations (\ref{gdiffeq1} - \ref{gdiffeq3}) immediately imply that $X_{\bar0}( F) =0 = X_{\bar0}( \bar F)$, which makes $F$ independent of $t$. Now we solve the remaining equation (\ref{gdiffeq4}) by imposing:
 $ \bar X_5 (F) \;=\; \bar X_6(F) \;=\; 0$ which read
\begin{equation}
     \frac{F_{\alpha}}{l} -i \frac{1}{l}( F_{\xi_1} \cot\alpha - F_{\xi_2}\tan\alpha - F_{\xi_3}\tan\alpha) =0,
\end{equation}
\begin{equation}
   \frac{F_{\beta}}{l} - i  \frac{1}{l}(F_{\xi_2}\cot\beta - F_{\xi_3}\tan\beta) = 0 \, .
\end{equation}
Substituting the ansatz \begin{eqnarray}
     F = \sum_{m,n,q} C_{mnq} (\alpha, \beta) e^{-i m\xi_1 - i n \xi_2- i q \xi_3}
 \end{eqnarray}
and solving the resulting equations gives (for constant $c_{mnq}$) 
 \begin{eqnarray}
   F = \sum_{m, n, q} c_{mnq} (\sin\alpha e^{-i \xi_1})^m (\sin\beta\cos\alpha e^{-i \xi_2})^n (\cos\alpha\cos\beta e^{-i \xi_3})^q.\label{giant2}
\end{eqnarray}
Therefore, in terms of the complex coordinates\begin{eqnarray}
      Z_1 = \sin\alpha e^{-i \xi_1},\;\;\;\;\; Z_2 = \sin\beta\cos\alpha e^{-i \xi_2},\;\;\;\;\;\; Z_3 = \cos\alpha\cos\beta e^{-i \xi_3}, \label{Zcoordinates}
  \end{eqnarray}
the general giants are given by $F(Z_1,\;Z_2,\;Z_3) = 0$. Just as before imposing $ \bar X_5 (\bar F) \;=\; \bar X_6(\bar F) = 0$ 
give completely equivalent class.  Remarkably, these giants are described by the same complex functions as in the $\omega = 0$ case of Mikhailov \cite{Mikhailov:2000ya}.


\section{The Kim-Lee type solutions}\label{kimleesection}
Kim and Lee \cite{Kim:2006he} provided a unified description of giants and dual-giants in $AdS_5 \times S^5$ in terms of D3-branes with world-volumes given by the common zeros of three independent holomorphic functions with specific homogeneity conditions. Here, for completeness, we provide a generalisation of these to the Gutowski-Reall black hole context.  

We postulate that here too the D3 configurations of interest are given by zeros of three complex functions $F^{(I)}$ (for $I=1,2,3$). Then similar to \cite{Ashok:2008fa}, we see that each of these functions has to satisfy vanishing conditions of the coefficients of $ \mathbi{E}^{\bar 1}, \mathbi{E}^{\bar 2},\;\mathbi{E}^{\bar 5},\;\mathbi{E}^{\bar6}$ and  $\mathbi{E}^{\bar 0}$ in (\ref{general constraint}). So, each of these general functions $F^{(I)}(r,\;t,\;\theta,\;\phi,\;\psi,\;\alpha,\;\beta,\;\xi_i)$ should satisfy five differential equations. The first of these is 
 $F^{(I)}_t = 0$, which makes $F^{(I)}$ independent of $t$. Then we use the following ansatz:
\begin{eqnarray}
    F^{(I)}= \sum_{n_1, n_2, m_1, m_2, m_3} C_{n_1, n_2, m_1, m_2, m_3}(r, \theta, \alpha, \beta) e^{i n_1 \phi + i n_2 \psi -i m_1 \xi_1  -i m_2 \xi_2 -i m_3 \xi_3  }.
\end{eqnarray}
Then the solutions to the remaining four equations can be obtained easily, and we find $$F^{(I)} = F^{(I)}(\Phi_1, \Phi_2, Z_1, Z_2, Z_3)$$ where
 \begin{eqnarray}
 \Phi_1 &=& \cot\frac{\theta}{2} e^{i\psi},  ~~~  \Phi_3 =  \left(\frac{r^2-\omega^2}{l^2+r^2+2\omega^2}\right)^{\frac{l^2}{l^2+3\omega^2}} \, \sin\theta \, e^{i\phi}\, ,\cr
 Z_i &=& (r^2-\omega^2)^\frac{\omega^2}{l^2+3\omega^2} (l^2+r^2+2\omega^2)^\frac{l^2+\omega^2}{2(l^2+3\omega^2)} \, \mu_i \, e^{-i\xi_i} \, .
 \end{eqnarray}
 where $\mu_1 = \sin\alpha$, $\mu_2 = \cos\alpha \, \sin\beta$ and $\mu_3 = \cos\alpha \, \cos\beta$ as in (\ref{Zcoordinates}).

From this way of describing the D3-branes, one can recover the giants and dual-giants we found in previous sections as special cases. For instance, if we take two of the $F^{(I)}$ to be $Z_1/Z_2-c_1$ and $Z_1/Z_3 - c_2$, leads to  $\alpha,\; \beta,\; \xi_{13}, \; \xi_{12}$ being constants. Then the final function being $F(\Phi_1, \Phi_3, (Z_1Z_2Z_3)^{1/3})$ corresponds to our dual-giants of section \ref{dualgiantsection}. Similarly, if we take two of the functions $F^{(I)}$ to be $\Phi_1-d_1$ and $\Phi_3-d_2$ and the third one to be $F(Z_1, Z_2, Z_3)$ corresponds to the giants of section \ref{giantsection}. 

Even though the geometric meaning of the complex variables $\Phi_a$ and $Z_i$ appear mysterious, we point out that all of them are solutions to the scalar Laplace equations in the 10d GR black hole geometry. This is a fact they share with their $\omega \rightarrow 0$ cousins of the pure $AdS_5 \times S^5$ giants.

Finally, it is easy to check that when we take $\omega \rightarrow 0$ this description maps to the one by \cite{Kim:2006he} (after appropriate identification of coordinates). 

\section{Turning on world-volume flux}\label{emwavesection}

In the previous sections we have found all the Mikhailov giants and the wobbling dual-giants in the GR black hole background. Here extend the analysis to include non-zero  world-volume electromagnetic fluxes on them that continue to preserve their supersymmetries. When the field strength $\mathbi{F}$ of the world-volume gauge field is turned on, the kappa projection condition becomes
\begin{eqnarray}
    \frac{\epsilon^{ijkl}}{\sqrt{-\rm{det}(h+\mathbi{F})}} \left[ \frac{1}{4!}\gamma_{ijkl}\epsilon+\frac{1}{4}\mathbi{F}_{ij}\gamma_{kl}\epsilon^* +\frac{1}{8}\mathbi{F}_{ij}\mathbi{F}_{kl}\epsilon\right] = \pm i \epsilon.
\end{eqnarray}
Since the brane should preserve the same supersymmetries as it does in the absence of the gauge field strength $\mathbi{F}$, the following conditions must be satisfied along with the conditions (\ref{fourgamma}),\;(\ref{twogamma}) and (\ref{timelike-cond}).
\begin{eqnarray}
    \mathbi{F} \;\wedge \;\mathbi{E}^{AB} &=& 0\nonumber\\
     \mathbi{F} \;\wedge \;\mathbi{E}^{\bar A\bar B} &=& 0\nonumber\\
    \mathbi{F}\;\wedge\; \mathbi{E}^{A\bar A} &=& 0\rm{~~~~for~~~~} A,B \;=\;{0,\;1,\;2,\;5,\;6}\nonumber\\
    \mathbi{F}\;\wedge\;\mathbi{F}&=& 0.
\end{eqnarray}
 Following the steps of \cite{Ashok:2010jv} and assuming that the gauge field strength is the pull-back of a spacetime 2-form onto the world-volume, the field strength can be written as
 \begin{eqnarray}
     \mathbi{F} = \rm{Re}(\chi_{01}\mathbi{E}^{01} \;+\;\chi_{02}\mathbi{E}^{02} \;+\; \chi_{12}\mathbi{E}^{12}).\label{F part}
 \end{eqnarray}
Here $\chi_{01},\;\chi_{02},\;\chi_{12}$ are arbitrary complex functions of spacetime coordinates restricted to the D3-brane world-volume. Next one imposes the Bianchi identity and the equation of motion
\begin{eqnarray}
    d\mathbi{F}\;=\;0 \;\;\;\;\;\; \rm{and} \;\;\;\;\;\;\; d\mathbi{X} \;=\; 0 \label{bianchi and eom}
\end{eqnarray}
where the 2-form $\mathbi{X}$ is defined as \begin{equation*}
    \mathbi{X}\;=\; \frac{1}{8} \epsilon_{ijkl}\sqrt{-\rm{det}(h+\mathbi{F})} [(h+\mathbi{F})^{-1}-(h-\mathbi{F})^{-1}]^{kl} \; d\sigma^i \;\wedge\; d\sigma^j\;.
\end{equation*}
 Using the expression (\ref{F part}) and the BPS conditions, one can simplify the above expression of $\mathbi{X}$ to
 \begin{eqnarray}
     \mathbi{X} \;=\; Im(\chi_{01}\mathbi{E}^{01} \;+\;\chi_{02}\mathbi{E}^{02} \;+\; \chi_{12}\mathbi{E}^{12}).
 \end{eqnarray}
Thus the two equations of (\ref{bianchi and eom}) can be combined into one for a complex 2-form ${\cal G}$ as
\begin{eqnarray}
    d{\cal G} :=\;d(\mathbi{F}+i\mathbi{X}) \;=\;0\;\;\;\;\;\;\;\;\rm{where}\;\;\;~~ {\cal G} \;=\; \chi_{01}\mathbi{E}^{01} \;+\;\chi_{02}\mathbi{E}^{02} \;+\; \chi_{12}\mathbi{E}^{12}.
\end{eqnarray}
Now on, we will focus on the dual-giant case for illustrative purposes, and rewrite the expression of ${\cal G}$ in terms of the complex coordinates $\Psi_0, \Psi_1, \Psi_2$ defined in (\ref{new complex coordinates}) as:
 \begin{eqnarray}
    {\cal G}\;:=\; {\cal G}_{01} \frac{d\Psi_0}{\Psi_0}\;\wedge\; \frac{d\Psi_1}{\Psi_1}+{\cal G}_{02} \frac{d\Psi_0}{\Psi_0}\;\wedge\; \frac{d\Psi_2}{\Psi_2}+{\cal G}_{12} \frac{d\Psi_1}{\Psi_1}\;\wedge\; \frac{d\Psi_2}{\Psi_2}
\end{eqnarray} 
using the following dictionary
\begin{eqnarray}
    \mathbi{E}^{01} &= &\sqrt{\frac{l^2+3\omega^2}{2}}\;\cosh\rho \sqrt{l^2+\omega^2-(l^2+3\omega^2)\cosh2\rho}\left(\cos^2\frac{\theta}{2}\;\frac{d\Psi_0}{\Psi_0}\wedge \frac{d\Psi_1}{\Psi_1} + \right.\nonumber\\
    &&\left.\sin^2\frac{\theta}{2}\;\frac{d\Psi_0}{\Psi_0} \wedge \frac{d\Psi_2}{\Psi_2}\right)\nonumber\\
    \mathbi{E}^{02}&=& i\frac{e^{i\phi}}{4\sqrt{2}l}\sqrt{l^2+\omega^2-(l^2+3\omega^2)\cosh2\rho}\;\sin\theta\;\left[ (l^2+\omega^2+(l^2+3\omega^2)\cosh2\rho)\left(\frac{d\Psi_0}{\Psi_0}\wedge \frac{d\Psi_1}{\Psi_1}\right.\right.\nonumber\\ &&\left.\left.- \frac{d\Psi_0}{\Psi_0} \wedge \frac{d\Psi_2}{\Psi_2}\right)+(l^2-\omega^2-(l^2+3\omega^2)\cosh2\rho) \frac{d\Psi_1}{\Psi_1} \wedge \frac{d\Psi_2}{\Psi_2}\right]\nonumber\\
    \mathbi{E}^{12}&=& i\frac{e^{i\phi}\sqrt{l^2+3\omega^2}}{4l}\cosh\rho\;\sin\theta\;(l^2+\omega^2-(l^2+3\omega^2)\cosh2\rho)\left(\frac{d\Psi_0}{\Psi_0}\!\wedge\! \frac{d\Psi_1}{\Psi_1}-\right.\nonumber\\ &&\left.\frac{d\Psi_0}{\Psi_0}\! \wedge\!\frac{d\Psi_2}{\Psi_2}- \frac{d\Psi_1}{\Psi_1}\!\wedge\! \frac{d\Psi_2}{\Psi_2}\right)
\end{eqnarray}
It is now easy to compute $d{\cal G}$ since the 2-forms 
$\frac{d\Psi_0}{\Psi_0}\wedge \frac{d\Psi_1}{\Psi_1}, \frac{d\Psi_0}{\Psi_0}\wedge \frac{d\Psi_2}{\Psi_2},\frac{d\Psi_1}{\Psi_1}\wedge\frac{d\Psi_2}{\Psi_2}$ have vanishing exterior derivatives. Finally, we obtain, from $d{\cal G}=0$
\begin{eqnarray}
   && d{\cal G}_{01} \frac{d\Psi_0}{\Psi_0}\;\wedge\; \frac{d\Psi_1}{\Psi_1}+d{\cal G}_{02} \frac{d\Psi_0}{\Psi_0}\;\wedge\; \frac{d\Psi_2}{\Psi_2}+d{\cal G}_{12} \frac{d\Psi_1}{\Psi_1}\;\wedge\; \frac{d\Psi_2}{\Psi_2}\;=\;0 ~~ {\rm where} \nonumber\\
    &&d{\cal G}_{ij}\;=\; X_1({\cal G}_{ij}) \mathbi{E}^1+ \bar{X_1}({\cal G}_{ij})  \bar{\mathbi{E}^1}+ X_2({\cal G}_{ij}) \mathbi{E}^2+ \bar{X_2}({\cal G}_{ij})  \bar{\mathbi{E}^2}+ X_0({\cal G}_{ij}) \mathbi{E}^0+ X_{\bar0}({\cal G}_{ij})  \mathbi{E}^{\bar0}.\nonumber\\
\end{eqnarray}
Here $X_i$ are the same differential operators as in (\ref{diffopsX}). Using the fact that the world-volume is given by the zeros of the holomorphic function $F(\Psi_i)$ we have $ a_0 \mathbi{E}^0+a_1\mathbi{E}^1+a_2\mathbi{E}^2=0$, which in turn implies
\begin{equation*}
    \mathbi{E}^{012}\;=\;\mathbi{E}^{0\bar1\bar2}\;=\;0 \, .
\end{equation*}
Since the other 3-forms are non-zero, the only possibility for solving $d{\cal G}=0$ is 
\begin{eqnarray}
    X_{\bar0}({\cal G}_{ij})\;=\; \bar X_{1}({\cal G}_{ij})\;=\; \bar X_{2}({\cal G}_{ij})\;=\;0.
\end{eqnarray}
These are similar equations to the ones we have already solved to obtain the embedding function $F$ for dual-giants, which immediately implies that ${\cal G}_{ij}$ are holomorphic functions of $(\Psi_0,\Psi_1,\Psi_2)$. Therefore, the gauge fields that preserves the same supersymmetries as the dual-giants and the GR black hole background is given by the real part of
 \begin{eqnarray}
     {\cal G}\;=\;&&\left[{\cal G}_{01}(\Psi_0, \Psi_1, \Psi_2) \frac{d\Psi_0}{\Psi_0}\;\wedge\; \frac{d\Psi_1}{\Psi_1}+{\cal G}_{02}(\Psi_0, \Psi_1, \Psi_2) \frac{d\Psi_0}{\Psi_0}\;\wedge\; \frac{d\Psi_2}{\Psi_2}\right.\nonumber\\&&~~~~~~~~~~~~~~~~~~~~~~~~~~~~~~~~~~~~~ \left.+\;{\cal G}_{12}(\Psi_0, \Psi_1, \Psi_2) \frac{d\Psi_1}{\Psi_1}\;\wedge\; \frac{d\Psi_2}{\Psi_2}\right].
 \end{eqnarray}
As expected, when we take $\omega \rightarrow 0$ limit of this answer we recover the ones found by \cite{Ashok:2010jv} in the case of pure $AdS_5 \times S^5$.

As a further illustration, let us compute the BPS gauge field on a $SU(2)_R$ invariant dual-giant described by the following six constraints.
 \begin{eqnarray}
    \rho\;=\; \rho_0,\;\;\;\;\;\;\; \xi_i\;=\;\xi_i^{(0)}, \;\;\;\;\;\;\;\;\alpha\;=\;\alpha_0, \;\;\;\;\;\;\;\beta\;=\;\beta_0\nonumber
\end{eqnarray}
where $i=1,2,3$, and the world-volume coordinates are $ t =\sigma_0,\;\theta=\sigma_1,\;\phi=\sigma_2,\;\psi=\sigma_3$. In this case the relevant 2-forms are given by
\begin{eqnarray}
    \mathbi{E}^{01}\;&=&\;0\nonumber\\
    \mathbi{E}^{02}&=& \frac{e^{i\sigma_2}}{4l} \sqrt{\omega^2+(l^2+3\omega^2)\sinh^2\rho_0}(2\omega^2+(l^2+3\omega^2)\sinh^2\rho_0)\nonumber\\
    && \times (i d\sigma_2\wedge d\sigma_1+\sin\sigma_1 d\sigma_2\wedge d\sigma_3+i \cos\sigma_1 d\sigma_3\wedge d\sigma_1)\nonumber\\
    \mathbi{E}^{12}&=& \frac{e^{i\sigma_2}}{4l} \sqrt{(l^2+3\omega^2)}\cosh\rho_0(\omega^2+(l^2+3\omega^2)\sinh^2\rho_0) \nonumber\\
    &&\times (d\sigma_2\wedge d\sigma_1-i\sin\sigma_1 d\sigma_2\wedge d\sigma_3+\cos\sigma_1 d\sigma_3\wedge d\sigma_1).
\end{eqnarray}
 Substituting these in the expression of ${\cal G}$, we find
 \begin{eqnarray}
     {\cal G}\;=\;-\frac{1}{2}{\cal G}_{12}(\Psi_0 = c,\;\Psi_1,\;\Psi_2) \;(i \csc\sigma_1 \, d\sigma_2 \wedge d\sigma_1 + d\sigma_2\wedge d\sigma_3+i\cot\sigma_1 \, d\sigma_3 \wedge d\sigma_1) \cr &&
 \end{eqnarray}
where
 \begin{equation*}{\cal G}_{12}(\Psi_0 = c,\Psi_1,\Psi_2) \;=\; \sum_{mn}C_{mn}\; \Psi_1^m \Psi_2^n \, .\end{equation*}
Thus we have obtained the BPS electromagnetic fields on the round dual-giants explicitly, which again match with the ones found in \cite{Sinha:2007ni, Ashok:2010jv} when $\omega=0$.

One can obtain the EM waves on the giant gravitons of previous section too in the GR black hole, whose details we omit.

\section{Discussion}\label{discussion}

We have presented the complete sets of BPS D3-brane configurations in the Gutowski-Reall black hole of type IIB string theory, that are generalisations of the well-studied Mikhailove giants \cite{Mikhailov:2000ya} and the wobbling dual-giants \cite{Ashok:2008fa} of pure $AdS_5 \times S^5$. Remarkably both the dual-giants and giants are given by zeros of holomorphic functions of appropriate complex combinations of the coordinates. In the case of the dual-giants the complex variables $\Psi_i$ ($i=0,1,2$) involved are non-trivial generalisations of those in $AdS_5 \times S^5$, whereas for the giants the $Z_i$ ($i=1,2,3$) are exactly the same as those in the context of pure the $AdS_5 \times S^5$. We have provided their Kim-Lee type description in a unified manner as simultaneous zeros of three independent complex functions of five complex combinations of coordinates all of which are scalar harmonics of the GR geometry. Our solutions spaces include the known exact dual-giant solutions in these black hole background as special case.\footnote{The embedding functions that describe wobbling dual-giants of \cite{Ashok:2008fa} also contain D3-branes that end of the boundary as BPS strings as shown in \cite{Ashok:2020hmq}. The dual-giant solutions we have found here also contain such configurations, and these probes could help compute some interesting properties of the GR black holes, which is worth pursuing.}

We have also shown how to turn on world-volume abelian gauge fields without breaking supersymmetries on BPS branes in the Gutowski-Reall black hole, which were not known earlier. Furthermore, we have also been able to find susy giants and dual-giants in the most general (extremal) 1/16-BPS black hole known in $AdS_5 \times S^5$ with two independent angular momenta in $AdS_5$ and one R-charge \cite{Aharony:2021zkr} (see Appendix \ref{dgingengr} for details). Finally, we have also shown that in the context of the only known smooth 1/16-BPS 1-parameter deformation of $AdS_5 \times S^5$ \cite{Gauntlett:2004cm}, the giants and dual-giants actually do not depend on the deformation parameter (see Appendix \ref{deformedgiants} for details).

The simpler set of dual-giants corresponding to $F(\Psi_0) = 0$ have played an interesting role in the context of end-points of instabilities of the non-extremal versions of the GR black holes \cite{Choi:2024xnv}. We have explicitly constructed the BPS electromagnetic fields on these probes, which when included should give more interesting back reactions on the black hole and could provide further understanding in the context of \cite{Choi:2024xnv}.

Even though we have found all solutions of dual-giants that are continuously connected to the ones in \cite{Aharony:2021zkr} (see Appendix \ref{appendix1} for details), the equations we obtained could potentially admit more solutions. We offer some comments in this regard in appendix \ref{moresolns}. It is yet unclear to us if any of the additional configurations we demonstrate there correspond to any physically viable finite-energy configurations of D3-branes or not. It will be important to settle this question -- as this could have implications even in the pure $AdS_5 \times S^5$ context.

A natural next step is characterisation and quantisation of the solutions spaces found here, and the counting of the BPS states (see, \cite{Biswas:2006tj, Mandal:2006tk, Ashok:2008fa} for some works for higher supersymmetric D3-brane solutions in $AdS_5 \times S^5$). One expects these steps to play important role in further understanding the 1/16-BPS state counting of the dual ${\cal N}=4$ SYM. We leave this analysis for future.

\section*{Acknowledgements}
We thank Gautam Mandal for posing a question to one of us which inspired in part investigations of this work. We are also grateful to Sujay Ashok for valuable discussions.

\appendix
\section{Perturbative solution to (\ref{general eq1})}
\label{appendix1}
In this appendix, we will show that  the holomorphic constraint (\ref{Fconstraint}) captures all the BPS dual-giants which are smooth deformations about the $F(\Psi_0) = 0$ dual-giants (the $SU(2)_R$ invariant dual-giants of \cite{Aharony:2021zkr}). For this we solve the equation (\ref{general eq1}) for $\rho$ and $\xi$ with the following perturbative ansatz:
\begin{eqnarray}
    \rho \;&=&\; \rho_0 \;+\; \epsilon\; \rho_1 (\theta,\;\phi,\;\psi)\; + \;\epsilon^2\; \rho_2 (\theta,\;\phi,\;\psi)\;+ \cdots\nonumber\\
    \xi \;&=&\; \xi_0 \;+\; \epsilon\; \xi_1 (\theta,\;\phi,\;\psi)\; + \;\epsilon^2\; \xi_2 (\theta,\;\phi,\;\psi)\;+ \cdots.
\end{eqnarray}
 Here the seed dual-giant, in the static gauge, is described by $\rho\;=\; \rho_0$ and $ \xi\;=\; \xi_0$. To the first order in $\epsilon$, equation (\ref{general eq1}) can be split into real and imaginary parts as
   \begin{eqnarray}
     \sin\theta\; \frac{\partial\tilde{\xi_1}}{\partial\theta}\;+\; \frac{\partial\tilde{\rho_1}}{\partial\psi}\;-\;\cos\theta\;\frac{\partial\tilde{\rho_1}}{\partial\phi}\;&=& 0\, ,\nonumber\\
     -\sin\theta\; \frac{\partial\tilde{\rho_1}}{\partial\theta}\;+\; \frac{\partial\tilde{\xi_1}}{\partial\psi}\;-\;\cos\theta\;\frac{\partial\tilde{\xi_1}}{\partial\phi}\;&=& 0.
 \end{eqnarray}
  Here, $\tilde{\xi_1}= (l^2+3\omega^2)\cosh\rho_0\sinh\rho_0 \;\xi_1$ and $ \tilde{\rho_1}= ((l^2+3\omega^2)\sinh^2\rho_0 + 2 \omega^2) \;\rho_1$. The solution to the above equations is
  \begin{eqnarray}
      \rho_1 \!\!&=& \!\!\sum_{m,\;n} (c^{11}_{mn}(\rho_0,\,\xi_0)\;\sin(m\phi\;+\;n\psi) + c^{12}_{mn}(\rho_0,\,\xi_0) \cos(m\phi + n\psi)) \left(\sin\tfrac{\theta}{2}\right)^{m-n} \!\!\! \left(\cos\tfrac{\theta}{2}\right)^{m+n}\nonumber\\
       \xi_1 \!\!&=& \!\!\sum_{m,n} (d^{11}_{mn}(\rho_0,\,\xi_0) \sin(m\phi + n\psi) + d^{12}_{mn}(\rho_0,\,\xi_0) \cos(m\phi+n\psi)) \left(\sin\tfrac{\theta}{2}\right)^{m-n} \left(\cos\tfrac{\theta}{2}\right)^{m+n}.\nonumber\label{rho1xi1}\\
  \end{eqnarray}
Similarly, one can work to higher orders systematically, and, in principle, solve (\ref{general eq1}) for $\rho$ and $\xi$ to arbitrary higher orders in $\epsilon$ expansion. 

We want to show that the holomorphic constraint $G(\Psi_0, \Psi_1, \Psi_2)=0$ contains these above dual-giant configurations. For this, one needs to expand the constants $C_{mnq}$  in 
\[G(\Psi_0,\;\Psi_1,\;\Psi_2) = \sum_{m,\;n,\;q} C_{mnq} \Psi_0^m\;\Psi_1^n\;\Psi_2^q\] 
as follows
\begin{eqnarray}
\label{cexps}
    C_{m00}\;&=&\; C^{(0)}_m \;+\; \epsilon\; C^{(1)}_m\;+\; \epsilon^2 \;C^{(2)}_m\;+\;\cdots\nonumber\\
    C_{mnq}\;&=&\; \epsilon\; C^{(1)}_{mnq}\;+\; \epsilon^2\; C^{(2)}_{mnq}\;+\;\cdots.
    \end{eqnarray}
where, $n,\; q\;\ne\, 0$. Substituting these expansions (\ref{cexps}) into $G$ one can expand it as
\begin{eqnarray}
    G = \sum_{m} C_m^{(0)} \;\Psi_0^m + \epsilon \sum_{m,\;n,\;q} C^{(1)}_{mnq}\Psi_0^m\;\Psi_1^n\;\Psi_2^q + \epsilon^2 \sum_{m,\;n,\;q} C^{(2)}_{mnq}\Psi_0^m\;\Psi_1^n\;\Psi_2^q +\cdots.
\end{eqnarray}
At the leading order the constraint $G=0$ describes only the seed dual-giant, i.e. $\rho=\rho_0,\; \xi=\xi_0$. To find the solution to ${\cal O}(\epsilon)$ we need to solve \begin{eqnarray}
    \sum_{m} C_m^{(0)} \;\Psi_0^m\;+\; \epsilon \; \sum_{m,\;n,\;q} C^{(1)}_{mnq}\Psi_0^m\;\Psi_1^n\;\Psi_2^q \;=\; 0.
\end{eqnarray}
One can solve this for $(\rho, \xi)$ to ${\cal O}(\epsilon)$ by substituting $\rho=\rho_0\;+\; \epsilon\; \rho_1$ and $\xi=\xi_0\;+\; \epsilon\; \xi_1 $ in the above constraint (and keeping terms to ${\cal O}(\epsilon)$) and  solving it for $\rho_1$ and $\xi_1$. These solutions thus obtained can be seen easily to be the same as we found in (\ref{rho1xi1}) after appropriate identifications of the constants involved. 

The equations (\ref{general eq1}) can easily be solved to ${\cal O}(\epsilon^2)$. Similarly one can take the constraint $G=0$ up to ${\cal O}(\epsilon^2)$, then substitute the expansion of $\rho$ and $\xi$ up to second order in $\epsilon$ and  solve it for $\rho_2$ and $\xi_2$. It can again be recast to match with the solutions to (\ref{general eq1}) to ${\cal O}(\epsilon^2)$. We have verified this process to ${\cal O}(\epsilon^5)$ successfully with Mathematica. This matching clearly indicates that the holomorphic constraint contains all the dual-giant configurations connected to the round ones through smooth deformations.


\section{ Dual-giants in more general GR black holes}
\label{dgingengr}
Even though we considered susy D3-branes in the Gutowski-Reall black hole, our methods can be used to achieve this exercise in the more general such black holes. To demonstrate the power of our methods, in this appendix we will provide all wobbling dual-giant type D3-brane probes in the generalisation of GR black hole to non-equal angular momentum. The first five vielbeins for such an $AdS_5$ black hole can be given in the orthotoric coordinate system as \cite{Aharony:2021zkr},
 \begin{eqnarray}
     e^0 &=& f(dt - w),\nonumber\\
     e^1 &=& \frac{1}{f^{1/2}}\sqrt{\frac{\eta-\xi}{\mathcal{F}(\xi)}} d\xi,\;\;\;\;\;\;\;\;\;\;e^2 = \frac{1}{f^{1/2}}\sqrt{\frac{\mathcal{F}(\xi)}{\eta-\xi}} (d\Phi +\eta d\Psi),\nonumber\\
     e^3&=&- \frac{1}{f^{1/2}}\sqrt{\frac{\eta-\xi}{\mathcal{G}(\xi)}} d\eta,\;\;\;\;\;\;\;\;\;\;e^4 = \frac{1}{f^{1/2}}\sqrt{\frac{\mathcal{G}(\xi)}{\eta-\xi}} (d\Phi +\eta d\Psi).
 \end{eqnarray}
where the functions involved are
\begin{eqnarray}
    \mathcal{G}(\eta) &=& - \frac{4(1-\eta^2)}{(a^2-b^2)\tilde{m}}\left[(1-a^2)(1+\eta)+(1-b^2)(1-\eta)\right]\equiv (\eta -g_1)(\eta-g_2)(\eta-g_3),\nonumber\\
    \mathcal{F}(\xi)&=&-\mathcal{G}(\xi)-\frac{4(1+\tilde{m}}{\tilde{m}}\left( \frac{2+a+b}{a-b}+\xi\right)^3\equiv(\xi-f_1)(\xi-f_2)(\xi-f_3),\nonumber\\
    f &=& \frac{24(\eta-\xi)}{\mathcal{F''}+\mathcal{G''}},\;\;\;\;\;\;\;\;\; \tilde{m} = \frac{m}{(a+b)(1+a)(1+b)(1+a+b)}-1.
\end{eqnarray}
 Here the black hole parameters are $(m,\;a,\;b)$. The one-form $w$ and the gauge field $A$ can be found in \cite{Aharony:2021zkr}. Let us write these in a short hand notation as\begin{eqnarray}
     w = w_{\phi} d\Phi +w_{\psi} d\Psi,\;\;\;\;\;\;\;\; \;\; A= A_t dt + A_{\phi} d\Phi +A_{\psi}d\Psi.
 \end{eqnarray}
 The $S^5$ part this geometry can be written in the following frame
 \begin{eqnarray}
     && e^5 = d\rho_s,\;\;\;\;\;\;\;\;\;\;\;\; e^6 = \frac{1}{4} \sin(2\rho_s) (d\zeta_s -\cos(\theta_s)d\phi_s), \;\;\;\;\;\;\;\;
      e^7 = \frac{1}{2}\sin(\rho_s) d\theta_s,\nonumber\\ &&e^8 = \frac{1}{2}\sin\rho_s \sin\theta_s d\phi_s,\;\;\;\;\;\;\; e^9 = \frac{1}{3} (d\psi_s +3 e^6 \;\tan\rho_s -d\zeta_s +2 A).
 \end{eqnarray}
 The killing spinor $\epsilon$, found in \cite{Aharony:2021zkr}, of this background satisfies 
\begin{eqnarray}
     \Gamma^{09}\epsilon = \epsilon,\;\;\;\;\;\;\Gamma^{12}\epsilon = -i \epsilon,\;\;\;\;\;\;\;\;\;\Gamma^{34}\epsilon=\Gamma^{56}\epsilon=\Gamma^{78}\epsilon=i\epsilon.
 \end{eqnarray}
\subsection*{Dual-giant solutions:}
We use the following combination of pulled-back vielbeins to express the BPS constraints on the world-volume of the D3-branes (dual-giant type):
\begin{eqnarray}
     \mathbi{E}^1 = \mathfrak{e}^1 - i \mathfrak{e}^2, \,\,\,\,\,\,\,\,\,\,\,\,\,\,\,\, \mathbi{E}^3=\mathfrak{e}^3 + i\mathfrak{e}^4 \nonumber\\
      \mathbi{E}^5=\mathfrak{e}^5 + i \mathfrak{e}^6, \,\,\,\,\,\,\,\,\,\,\,\,\,\,\,\,\,  \mathbi{E}^7=\mathfrak{e}^7 + i\mathfrak{e}^8 \nonumber\\
      \mathbi{E}^0=\mathfrak{e}^0+\mathfrak{e}^9, \,\,\,\,\,\,\,\,\,\,\,\,\,\,\,\,\,\,  \mathbi{E}^{\bar 0}=\mathfrak{e}^0-\mathfrak{e}^9 .
\end{eqnarray}
After a straightforward analysis on the lines of section \ref{dualgiantsection} the corresponding BPS conditions turn out to be
 \begin{eqnarray}
     \mathbi{E}^{0\bar 0 13}\;= \;\mathbi{E}^{1\bar 1 3\bar 3}\;=\;\mathbi{E}^{013\bar 3}\;=\; \mathbi{E}^{031\bar1}\;=\;0
 \end{eqnarray}
along with their complex conjugates. We take the three complex constraints on the world volume of the D3-brane to be 
\begin{eqnarray}
    \mathbi{E}^5 \;=\;\mathbi{E}^7\;=\; 0 ~~~~~~{\rm and}~~~~~~~ J(t,\;\xi,\;\eta,\;\Phi,\;\Psi,\;\rho_s,\;\theta_s,\;\zeta_s,\;\phi_s,\;\psi_s) =0.
\end{eqnarray}
where $J$ is a complex function. The first two constraints will give the following
\begin{eqnarray}
    \rho_s =const.\;\;\;\;\;\; \theta_s =const.\;\;\;\;\;\; \zeta_s=const.\;\;\;\;\;\; \phi_s=const.
\end{eqnarray} 
 Using the BPS conditions, one can find equations similar to (\ref{diffeq1})-(\ref{diffeq4}). In this case, the vector fields are the following.
\begin{eqnarray}
    X_1 &=& \sqrt{\frac{f}{4(\eta-\xi)\mathcal{F}}}\left(\mathcal{F}\frac{\partial}{\partial\xi} + i(-2 A_{\psi}+2 A_{\phi} \xi +2 A_t w_{\phi} \xi -2A_t w_{\psi}) \frac{\partial}{\partial \psi_s}\right.\nonumber\\ & &\;\;\;\;\;\;\;\;\;\;\;\;\;\;\;\;\;\;\;\;\;\;\;\;\;\;\;\;\left. +i (w_{\psi}-w_{\phi} \xi) \frac{\partial}{\partial t} +i\frac{\partial}{\partial \Psi}-i \xi \frac{\partial}{\partial \Phi}\right)\nonumber\\
    X_3 &=& \sqrt{\frac{f}{4(\eta-\xi)\mathcal{G}}}\left(\mathcal{G}\frac{\partial}{\partial\eta} - i(-2 A_{\psi}+2 A_{\phi} \eta -2 A_t w_{\phi} \eta +2A_t w_{\psi}) \frac{\partial}{\partial \psi_s} \right.\nonumber\\&&\;\;\;\;\;\;\;\;\;\;\;\;\;\;\;\;\;\;\;\;\;\;\;\;\;\;\;\;\left.+i (w_{\psi}-w_{\phi} \eta) \frac{\partial}{\partial t} -i\frac{\partial}{\partial \Psi}+i \eta \frac{\partial}{\partial \Phi}\right)\nonumber\\
    X_0&=& \frac{1}{2 f}\left(\frac{\partial}{\partial t} -(2A_t - 3 f) \frac{\partial}{\partial \psi_s}\right)\;\;\;\;\;\;\;\;\;\;\;\;\;\;\nonumber\\
    X_{\bar0}&=& \frac{1}{2 f}\left(\frac{\partial}{\partial t} -(2A_t + 3 f) \frac{\partial}{\partial \psi_s}\right) \;=\; \frac{1}{2f} \left(\frac{\partial}{\partial t} -(3- 2 \alpha) \frac{\partial}{\partial \psi_s}\right).
\end{eqnarray}
From our previous analysis it can be shown that  all possible dual-giant type solutions can be found by solving the following:
\begin{eqnarray}
    X_{\bar0}(\bar J) X_0(J) - X_{\bar0} (J) X_0(\bar J)&=&0,
    \nonumber\\
    X_{\bar0}(\bar J) \bar X_1(J) - X_{\bar0} (J) \bar X_1(\bar J)&=& 0, \nonumber\\
    X_{\bar0}(\bar J) \bar X_3(J) - X_{\bar0} (J) \bar X_3(\bar J) &=& 0, \nonumber\\
    \bar{X_1}(J)\bar{X_3}(\bar J) - \bar{X_1}(\bar J) \bar{X_3}(J)&=& 0.\label{general equation for non equal angular momentum case}
\end{eqnarray}
The solutions in this case involve the roots of the polynomial $\mathcal{F}(\xi)$ and $\mathcal{G}(\eta)$ that make the expressions of $J$ much bigger and complicated. So we restrict to finding the dual-giants in the stable extremal black holes. The extremality limit should be taken by introducing the following scaling of the coordinates and the re-definition of $m$ and taking the scaling parameter $\lambda \rightarrow 0$.
\begin{eqnarray}
    m \;=\; (1+a)(1+b)(a+b)(1+a+b) +\lambda(\frac{1}{a-b}),&&\;\;\;\;\;\;\;\;\; \nonumber\\\xi \;=\; -\frac{\tilde{\xi}}{\lambda},\;\;\;\;\; \Phi\;=\; \lambda\tilde{\Phi},\;\;\;\;\;\;\Psi\;=\; \lambda\tilde{\Psi}&&
\end{eqnarray}
By taking this extremality limit and following the same steps as for the Gutowski-Reall black hole case, all the solutions of  equations (\ref{general equation for non equal angular momentum case}) can be found by taking  the general ansatz for the $J$ as\begin{eqnarray}
    J = \sum_{n_1,n_2,p,q} C_{n_1n_2pq}(\tilde{\xi}, \eta) \;e^{i n_1\tilde{\Phi}+i n_2 \tilde{\Psi}+i q \psi_s +i p t}.
\end{eqnarray}
Using this ansatz, finally we need to solve with the constraint $p\;=\; q(3-2\alpha)$.
\begin{eqnarray}
  && - F(\tilde{\xi}) \frac{\partial C_{n_1n_2pq}}{\partial \xi} +\tilde{\xi} (n_1 + p W_{\phi}-2q (a_{\phi}+a_t W_{\phi}))C_{n_1n_2pq}\;=\;0 \, ,\nonumber\\
 && G(\eta)\frac{\partial C_{n_1n_2pq}}{\partial \eta} \cr
 && +(2q a_{\psi}-n_2+n_1 \eta-2a_{\phi} q \eta + p \eta W_{\phi}-2a_t q\eta W_{\phi}-p W_{\psi}+2 a_t q W_{\psi}) C_{n_1n_2pq} = 0.\nonumber\\
\end{eqnarray}
The functions in the above equations are given by the following expansions,
\begin{eqnarray}
   && \mathcal{F} = \frac{F(\tilde{\xi})}{\lambda^3} + \mathcal{O}(\frac{1}{\lambda^2}),\;\;\;\;\;\; \mathcal{G} = \frac{G(\eta)}{\lambda} + \mathcal{O}(\lambda^0)\nonumber\\
  && \lambda \omega_{\phi}= W_{\phi} +\mathcal{O}(\lambda),\;\;\;\;\;\;\;\; \lambda\omega_{\psi}=W_{\psi} +\mathcal{O}(\lambda),\nonumber\\
  && \lambda A_{\phi}= a_{\phi} +\mathcal{O}(\lambda),\;\;\;\;\; \lambda A_{\psi}= a_{\psi} +\mathcal{O}(\lambda),\;\;\;\;\;\;\;\;  A_{t}= a_{t} +\mathcal{O}(\lambda).
\end{eqnarray}
After solving these equations we arrive at final solutions is given by 
\begin{small}
\begin{eqnarray}
   && J \;=\; \sum_{n_1,\;n_2,\;q} (1-\eta)^{\frac{n_1-n_2}{16(a-1)(1+a)^2(1+b)(1+a+b)}}(1+\eta)^{\frac{n_1+n_2}{16(1+a)(b-1)(1+b)^2(1+a+b)}} \nonumber\\&& \times (2+b^2(\eta-1)-a^2(1+\eta))^{-\frac{(a^2+b^2-2)n_1+(a-b)(a+b)n_2-24(a-1)(1+a)^2(b-1)(1+b)^2(1+a+b)q}{16(a-1)(1+a)^2(b-1)(1+b)^2(1+a+b)}}\nonumber\\
  &&  \times (2(1+a)(1+b)(1+a+b)(1+a^2+3a(1+b)+b(3+b))-\tilde{\xi})^{\frac{n_1+12(1+a)^2(1+b)^2(1+a+b)q}{8(1+a)(1+b)(1+a^2+3a(1+b)+b(3+b))(1+a+b)}}\nonumber\\
  && \times \;\tilde{\xi}^{-\frac{n_1-12(a+b)(2+a+b)(1+a)(1+b)(1+a+b)q}{8(1+a)(1+b)(1+a+b)(1+a^2+3a(1+b)+b(3+b))}}\times e^{in_1 \tilde{\Phi}+in_2\tilde{\Psi}+iq(\psi_s+(3-2\alpha)t)}.
\end{eqnarray}
\end{small}
Thus we have solved for the wobbling dual-giant configurations in this background as well, which demonstrates the utility of our methods. 

\section{Giants in the deformed $AdS_5 \times S^5$}
\label{deformedgiants}
 There exists a smooth 1-parameter deformation of $AdS_5 \times S^5$ background  \cite{Gauntlett:2004cm} that preserves two out of 32 supersymmetries of $AdS_5 \times S^5$. The frame for the deformed background is given by \begin{eqnarray}
    && e^0  = dt + \frac{r^2}{2 l}\sigma_3^L + \frac{f r^2}{1+\frac{r^2}{l^2}}\sigma_1^L,\;\;\;\;\;\;\;\;\;\; e^1 = \frac{dr}{1+\frac{r^2}{l^2}},\nonumber\\
    && e^2 = \frac{r}{2}\sigma_1^L, \;\;\;\;\;\;\;\;\;\;\;\;e^3 = \frac{r}{2}\sigma_2^L,\;\;\;\;\;\;\;\;\;\;e^4 = \frac{r\sqrt{1+\frac{r^2}{l^2}}}{2}\sigma_3^L,
 \end{eqnarray}
  and the rest of the vielbeins are  same as (\ref{gr-s5}) with $A = \frac{\sqrt{3}}{2}\frac{f r^2}{1+\frac{r^2}{l^2}}\sigma_1^L$. Here $f$ is the deformation parameter. The killing spinor $\epsilon$ of the background can be constraint by the following five projection conditions:
  \begin{eqnarray}
      \Gamma^{14} \epsilon = -i \epsilon,\;\;\;\;\;\;\;\; \Gamma^{23} \epsilon = i \epsilon,\;\;\;\;\;\;\;\; \Gamma^{57} \epsilon = -i \epsilon,\;\;\;\;\;\;\;\; \Gamma^{68} \epsilon = -i \epsilon,\;\;\;\;\;\;\;\; \Gamma^{09} \epsilon =  \epsilon. 
  \end{eqnarray}
In this case, we will use the following frame basis to write the BPS constraints
  \begin{eqnarray}
      \mathbi{E}^1 = \mathfrak{e}^1 - i \mathfrak{e}^4, \,\,\,\,\,\,\,\,\,\,\,\,\,\,\,\, \mathbi{E}^2=\mathfrak{e}^2 + i\mathfrak{e}^3 \nonumber\\
      \mathbi{E}^5=\mathfrak{e}^5 - i \mathfrak{e}^7, \,\,\,\,\,\,\,\,\,\,\,\,\,\,\,\,\,  \mathbi{E}^6=\mathfrak{e}^6 - i\mathfrak{e}^8 \nonumber\\
      \mathbi{E}^0=\mathfrak{e}^0+\mathfrak{e}^9, \,\,\,\,\,\,\,\,\,\,\,\,\,\,\,\,\,\,  \mathbi{E}^{\bar 0}=\mathfrak{e}^0-\mathfrak{e}^9 .
\end{eqnarray} 
In this notation the BPS conditions for the dual-giant type D3-branes workout to be  
\begin{eqnarray}
     \mathbi{E}^{0\bar 0 12}\;= \;\mathbi{E}^{1\bar 1 2\bar 2}\;=\;\mathbi{E}^{012\bar 2}\;=\; \mathbi{E}^{021\bar1}\;=\;0.
 \end{eqnarray}
Restricting to dual-giant type configurations we simply impose 
\begin{eqnarray}
     \mathbi{E}^5\;=\;\mathbi{E}^6\;=\;0.
\end{eqnarray}
Then the world-volume can be given by zeros of a single complex function $F$. Following the analysis of section \ref{dualgiantsection} we find that the dual-giants can be obtained by imposing 
\begin{eqnarray}
    X_{\bar 0}(F)\;=\; \bar{X_1}(F) \;=\; \bar{X_2}(F)\;=\; 0
 \end{eqnarray}
 where
\begin{eqnarray}
         X_1 &=&  \sinh\rho\cosh\rho\; \frac{\partial}{\partial \rho} -i\;l\sinh^2\rho  \frac{\partial}{\partial t} + 2i\; \frac{\partial}{\partial \phi},\cr
      X_2 &=& \frac{e^{i\phi}\csc\theta}{l \sinh\rho}\left(-i\;\sin\theta \frac{\partial}{\partial\theta} +\cos\theta\frac{\partial}{\partial\phi} -\frac{\partial}{\partial\psi}\right) - f\;\frac{\tanh\rho}{\cosh\rho} \left( l \frac{\partial}{\partial t} + \frac{\partial}{\partial \xi_1}+\frac{\partial}{\partial \xi_2}+\frac{\partial}{\partial \xi_3}\right),\cr
      X_0&=& \frac{1}{2 l}\left( l \frac{\partial}{\partial t} - \frac{\partial}{\partial \xi_1}-\frac{\partial}{\partial \xi_2}-\frac{\partial}{\partial \xi_3}\right),\;\;\;\;\;\;\;\; X_{\bar 0}=\frac{1}{2 l}\left( l \frac{\partial}{\partial t} + \frac{\partial}{\partial \xi_1}+\frac{\partial}{\partial \xi_2}+\frac{\partial}{\partial \xi_3}\right).
\end{eqnarray}
 Here we have used $r= l \, \sinh\rho$. It easily follows that the solution to these equations can be given in terms of holomorphic functions of the following coordinates.\begin{eqnarray}
     \psi_0 = l \cosh\rho\;e^{-i\xi},\; \psi_1= l \sinh\rho\;\cos\frac{\theta}{2}e^{-\frac{i}{2}(\phi+\psi+2\xi)}, \;\psi_2=l \sinh\rho\;\sin\frac{\theta}{2}e^{-\frac{i}{2}(\phi-\psi+2\xi)}.\cr
 \end{eqnarray}
Remarkably these are the same dual-giants that were found in \cite{Ashok:2008fa} for wobbling dual-giants of $AdS_5 \times S^5$. We can find the Mikhailov type giants too in this background with the same result as in section \ref{giantsection}.

\section{Any more BPS D3-branes?}
\label{moresolns}
As emphasised at the end of section \ref{dualgiantsection}, it is important to check if there are other physically viable solutions to our BPS equations. We initiate this exercise here in the context of dual-giants. Recall that in section \ref{dualgiantsection} we were left to solve the equation  $$\bar{X_1}(F)\bar{X_2}(\bar F) - \bar{X_1}(\bar F) \bar{X_2}(F)= 0$$ which can be rewritten as the singularity condition of the matrix 
\begin{eqnarray}
\left(\begin{array}{cc} \bar X_1(F) & \bar X_1(\bar F) \\ \bar X_2(F)  & \bar X_2(\bar F)  \end{array} \right).
\end{eqnarray}
One can impose this condition by taking either the two rows or two columns to be linearly independent with functional coefficients. This leads one to consider the two further sets of conditions
\begin{eqnarray}
   (i) \; &:&\;\; \bar X_1 (F) = \lambda_1(x^i)\;\bar X_2 (F) ~~\& ~~
 \bar X_1 (\bar F) = \lambda_1(x^i)\;\bar X_2 (\bar F) \label{eq63}\\
       (ii)\;& :&\;\; \bar X_1 (F) =\lambda_2(x^i)\; \bar X_1 (\bar F) ~~ \& ~~
 \bar X_2 (F) = \lambda_2(x^i)\;\bar X_2 (\bar F).\label{eq64}
\end{eqnarray}
where $\lambda_1(x^i)$ and $\lambda_2(x^i)$ can be any general complex functions of all spacetime coordinates $x^i$. Here we explore if these equations admit any more solutions that continue to satisfy the non-vanishing conditions (\ref{not equal conditions}).

The simplest way to solve the equations in (\ref{eq63}) or (\ref{eq64}) would have been to take $F = W(f(x^i))$, where $f$ is an arbitrary real function and $W$ is a complex functional. However, such a solution fails to satisfy the non-vanishing conditions and so we can discard them safely. However, for the purposes of demonstration, below we provide one sample solution to the equations of interest whose physical interpretation, if any, is unclear to us. 

The equation (\ref{diffeq4}) is same as the equation (\ref{complex vector field equation}) after using the coordinate transformation  (\ref{radial defination}). So the equations in (\ref{eq63}) and (\ref{eq64}) can be alternatively written in terms of the vector fields $X$ and $Y$ in (\ref{XYdefs}).
\begin{eqnarray}
  (i)\;\;  X(F)=\lambda(x^i) Y(F) ~~~ \& ~~~ X(\bar F) = \lambda(x^i) Y(\bar F)\label{xy}\\
  (ii)\;\;  X(F)=\delta(x^i) X(\bar F) ~~~ \& ~~~ Y( F) = \delta(x^i) Y(\bar F)\label{xx}.
\end{eqnarray}
Here $\lambda(x^i)$ and $\delta(x^i)$ both are non-vanishing complex functions of spacetime coordinates. First, we need to find what the possible choices $\lambda$ and $\delta$ are such that the equations (\ref{xy}, \ref{xx}) admit solutions. Below we will restrict to (\ref{xy}) and demonstrate that there are potentially other solutions to this equation. 
 

We first rewrite  (\ref{xy}) by taking combinations as
 \begin{eqnarray}\label{lambda F}
 &&  -i(2\omega^2+(l^2+3\omega^2)\sinh^2\rho)\frac{\partial F}{\partial\xi} +2 i l^2 \frac{\partial F}{\partial\phi} = -i \lambda_r \sin\theta\frac{\partial F}{\partial\theta}+i \;\lambda_i\left(\cos\theta\frac{\partial F}{\partial\phi}-\frac{\partial F}{\partial\psi}  \right)  \, , \cr
    && (l^2+3 \omega^2) \sinh\rho\;\cosh\rho\frac{\partial F}{\partial\rho}= \lambda_i \;\sin\theta\frac{\partial F}{\partial\theta}+ \;\lambda_r\left(\cos\theta\frac{\partial F}{\partial\phi}-\frac{\partial F}{\partial\psi} \right),
    \end{eqnarray}
where $\lambda_i$ and $\lambda_r$ are the real and imaginary parts of $\lambda(x^i)$. The solution for $F$ should be in the form of (\ref{trialf}). In general $\lambda(x^i)$ can also be written as
\begin{eqnarray}
    \lambda(x^i) = \sum_{m,\;n,\;q} D_{mnq}(\rho, \theta)e^{i m \phi+ i n \psi+ i q \xi}.
\end{eqnarray}
We further simplify the problem by restricting $\lambda$ to 
$\lambda(x^i) = \lambda(\rho, \theta) \equiv \lambda_r(\rho, \theta) + i \lambda_i (\rho, \theta)$.
By substituting these ansatz for $F$ and $\lambda$ into (\ref{lambda F}), we obtain
\begin{eqnarray}
\label{d8}
    && (l^2+3 \omega^2) \sinh\rho\;\cosh\rho\frac{\partial C_{mnq} }{\partial\rho}= \lambda_i \;\sin\theta\frac{\partial C_{mnq}}{\partial\theta}+ \; i \lambda_r\left(m\;\cos\theta-n \right)C_{mnq}(\rho,\;\theta),\nonumber\\
   &&  ((2\omega^2+(l^2+3\omega^2)\sinh^2\rho)q -2 l^2 m)C_{mnq}(\rho,\;\theta) = \cr && ~~~~~~~~~~~~~~~~~~~~~~~~~~~~~~~~~ -i \lambda_r \sin\theta\frac{\partial C_{mnq}}{\partial\theta}- \;\lambda_i\left(m\;\cos\theta-n  \right)C_{mnq}(\rho,\;\theta).
\end{eqnarray}
It turns out that if we set $ \lambda_r = 0$, for integrability of these equations allows for a non-vanishing $\lambda_i= \frac{2 l^2}{\cos\theta}$ only when $n=q=0$. But this forces the solution for $F$ to be $\xi$ independent, which in turn will make it fail to satisfy the non-vanishing conditions.
Thus we must have $ \lambda_r \ne 0$ which allows us to rewrite (\ref{d8}) as
\begin{eqnarray}
\label{two partial xy}
\frac{\partial \ln \, C_{mnq}}{\partial\rho}&=& i\left(\frac{\lambda_i((2\omega^2+(l^2+3\omega^2)\sinh^2\rho)q -2 l^2 m)}{\lambda_r (l^2+3 \omega^2) \sinh\rho\;\cosh\rho}+ \frac{(\lambda_i^2+\lambda_r^2)\left(m\;\cos\theta-n  \right)}{\lambda_r (l^2+3 \omega^2) \sinh\rho\;\cosh\rho} \right) \, , \cr
    \frac{\partial \ln \, C_{mnq}}{\partial\theta}&=& i\left(\frac{((2\omega^2+(l^2+3\omega^2)\sinh^2\rho)q -2 l^2 m)}{\lambda_r \sin\theta}\;+\; \frac{\lambda_i\left(m\;\cos\theta-n  \right)}{\lambda_r \sin\theta} \right) \, .
      \end{eqnarray}
Integrability of these partial differential equations will require $\lambda_i$ and $\lambda_r$ satisfy 

\begin{eqnarray}
\label{integrability}
  &&  \frac{\partial}{\partial\rho}\left(\frac{((2\omega^2+(l^2+3\omega^2)\sinh^2\rho)q -2 l^2 m)}{\lambda_r \sin\theta}\;+\; \frac{\lambda_i\left(m\;\cos\theta-n  \right)}{\lambda_r \sin\theta} \right)\\
    &=& \frac{\partial}{\partial \theta}\left(\frac{\lambda_i((2\omega^2+(l^2+3\omega^2)\sinh^2\rho)q -2 l^2 m)}{\lambda_r (l^2+3 \omega^2) \sinh\rho\;\cosh\rho}\;+\; \frac{(\lambda_i^2+\lambda_r^2)\left(m\;\cos\theta-n  \right)}{\lambda_r (l^2+3 \omega^2) \sinh\rho\;\cosh\rho} \right).\nonumber
\end{eqnarray}
This clearly eliminate constant $\lambda$. It is easy to check that when $(m,  n, q) \ne (0,0,0)$, there will be no solutions. But if $m = 0$ or $n =0$ the equation (\ref{two partial xy}) could and does admit solutions. For example 
 \begin{equation*}
    \lambda_r = \lambda_i = 2\omega^2+(l^2+3\omega^2)\sinh^2\rho
\end{equation*}
is a solution to (\ref{integrability}), using which one can solve (\ref{two partial xy}), and we find 
 \begin{eqnarray}
     F = \sum_{n,q} c_{nq} \, e^{i n \psi + i q \xi} e^{i(n-q)\log\tan\frac{\theta}{2} + (2n-q) \log((\sinh\rho)^{\frac{2\omega^2}{l^2 + 3 \omega^2}}(\cosh\rho)^{\frac{l^2 +\omega^2}{l^2 + 3 \omega^2}})}.
 \end{eqnarray}
This solution satisfies all the non-vanishing conditions, provided we consider at least two non-vanishing $c_{nq}$s in the sum, and is, a priori, not captured by the holomorphic class of solutions of section \ref{dualgiantsection}. 
We suspect that there could be many more solutions for general $\lambda(x^i)$ which are not in the  holomorphic class. However, if they are physically relevant or not is something we are still working on.

In the case of (\ref{xx}), one can show that  there are no solutions for $\delta (x^i) = \delta(\rho, \theta)$. But for more general $\delta$ there might have some interesting solutions. We leave this interesting analysis for future studies.

\providecommand{\href}[2]{#2}\begingroup\raggedright\endgroup

    \end{document}